\documentclass[sigconf, screen]{acmart}
\usepackage{lipsum}
\usepackage[english]{babel}
\usepackage{graphicx} 
\usepackage{spverbatim}
\usepackage[nolist]{acronym}
\usepackage[per-mode=fraction]{siunitx}
\usepackage{enumitem}
\usepackage{subcaption}
\usepackage{listings}

\usepackage{tabularx}
\usepackage{booktabs}
\usepackage{xcolor}
\usepackage{hyphenat}
\usepackage{pifont}
\newcommand{\cmark}{\textcolor{green!50!black}{\ding{51}}}
\newcommand{\xmark}{\textcolor{red!70!black}{\ding{55}}}

\usepackage{arydshln}

\AtBeginDocument{%
  }

\lstdefinestyle{distinct}{
    backgroundcolor=\color{gray!10}, 
    frame=single,                   
    rulecolor=\color{black},        
    basicstyle=\footnotesize\ttfamily, 
    breaklines=true,                
    xleftmargin=0.05\textwidth,      
    xrightmargin=0.05\textwidth,     
    captionpos=b                    
}

\copyrightyear{2026}
\acmYear{2026}
\setcopyright{cc}
\setcctype{by}
\acmConference[CHI '26]{Proceedings of the 2026 CHI Conference on Human Factors in Computing Systems}{April 13--17, 2026}{Barcelona, Spain}
\acmBooktitle{Proceedings of the 2026 CHI Conference on Human Factors in Computing Systems (CHI '26), April 13--17, 2026, Barcelona, Spain}
\acmPrice{}
\acmDOI{10.1145/3772318.3790772}
\acmISBN{979-8-4007-2278-3/2026/04}

\begin{document}

\title{GTA: Generative Traffic Agents for Simulating Realistic Mobility Behavior}


\author{Simon Lämmer}
\email{simon.laemmer@outlook.de}
\orcid{0009-0006-1144-3554}
\affiliation{%
	\institution{Leipzig University}
	\city{Leipzig}
	\country{Germany}}

\author{Mark Colley}
\orcid{0000-0001-5207-5029}
\email{m.colley@ucl.ac.uk}
\affiliation{%
  \institution{UCL Interaction Centre}
  \city{London}
  \country{United Kingdom}
}

\author{Patrick Ebel}
\email{ebel@uni-leipzig.de}
\orcid{0000-0002-4437-2821}
\affiliation{%
	\institution{ScaDS.AI, Leipzig University}
	\city{Leipzig}
	\country{Germany}}

\renewcommand{\shortauthors}{Lämmer, Colley, and Ebel}

\begin{abstract}
People's transportation choices reflect complex trade-offs shaped by personal preferences, social norms, and technology acceptance. Predicting such behavior at scale is a critical challenge with major implications for urban planning and sustainable transport. Traditional methods use handcrafted assumptions and costly data collection, making them impractical for early-stage evaluations of new technologies or policies. We introduce Generative Traffic Agents (GTA) for simulating large-scale, context-sensitive transportation choices using LLM-powered, persona-based agents. GTA generates artificial populations from census-based sociodemographic data. It simulates activity schedules and mode choices, enabling scalable, human-like simulations without handcrafted rules. We evaluate GTA in Berlin-scale experiments, comparing simulation results against empirical data. While agents replicate patterns, such as modal split by socioeconomic status, they show systematic biases in trip length and mode preference. GTA offers new opportunities for modeling how future innovations, from bike lanes to transit apps, shape mobility decisions.

\end{abstract}

\ccsdesc[500]{Human-centered computing~Interactive systems and tools}
\ccsdesc[500]{Computing methodologies~Natural language processing}

\keywords{agents, generative AI, large language models, multi-agent simulation, traffic simulation, public transportation}

\begin{teaserfigure}
\centering
  \includegraphics[width=\textwidth]{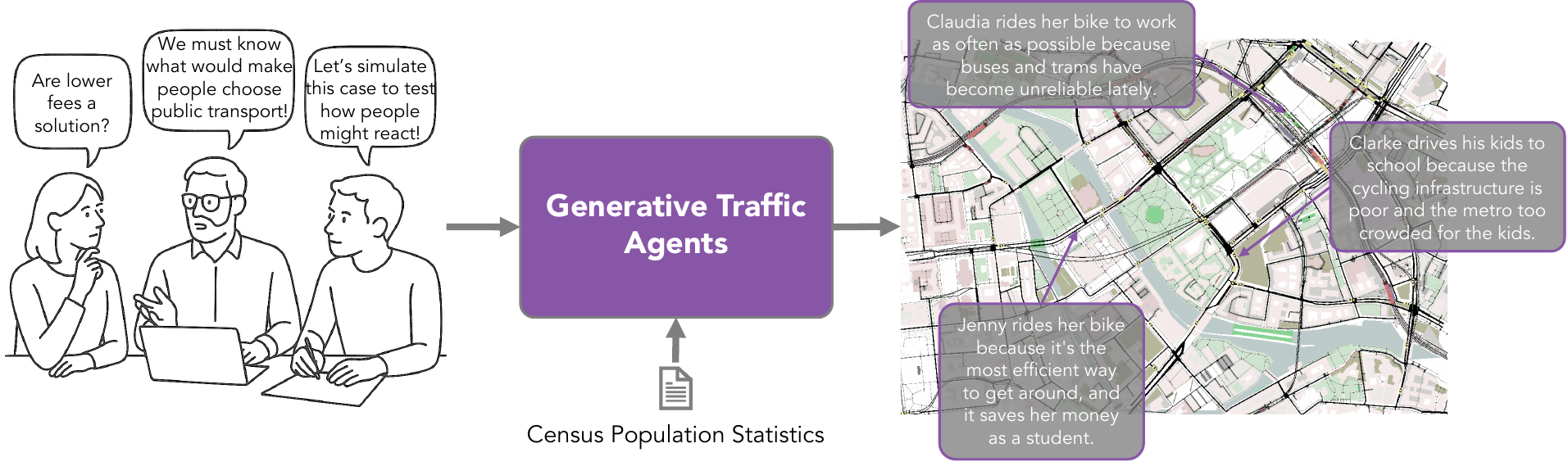}
  \caption{GTA enables traffic planners, policymakers, and product designers to conduct large-scale, agent-based traffic simulations based on population statistics. This allows them to gain insights in the decision-making processes that drive the adoption of mobility innovations.}
  \Description{A diagram showing how generative traffic agents, informed by census data, can simulate individual transportation decisions in a city. On the left, three people are shown discussing transport policy questions. In the center, a purple box labeled Generative Traffic Agents connects census population statistics to a city map on the right. On the right, a map of a city is overlaid with examples of individual travel decisions produced by the agents.}
  \label{fig:teaser}
\end{teaserfigure}

\hyphenation{anony-mized}

\maketitle

\begin{acronym}
  \acro{ACC}{Advanced Cruise Control}
  \acro{ACT-R}{Adaptive Control of Thought—Rational}
  \acro{CI}{Continuous Integration}
  \acro{CE}{Continuous Experimentation}
  \acro{UI}{User Interface}
  \acro{UX}{User Experience}
  \acro{UXD}{User Experience Design}
  \acro{XAI}{Explainable AI}
  \acro{GA}{Generative Agent}
  \acro{GTA}{Generative Traffic Agents}
  \acro{LLM}{Large Language Model}
  \acro{SUMO}{Simulation of Urban MObility}
  \acro{TraCI}{Traffic Control Interface}
  \acro{OSM}{OpenStreetMap}
  \acro{SR}{synthetic reconstruction}
  \acro{CO}{combinatorial optimization}
  \acro{SL}{statistical learning}
  \acro{ML}{Machine Learning}
  \acro{ANES}{American National Election Studies}
  \acro{CoT}{Chain of Thought}
  \acro{LM}{Language Modeling}
  \acro{VLM}{visual language models}
  \acro{ST-LLM}{Spatial-Temporal Large Language Model}
  \acro{CEMDAP}{Comprehensive Econometric Microsimulator for Daily Activity-Travel Patterns}
  \acro{Cadyts}{Calibration of Dynamic Traffic Simulations}
  \acro{IPF}{Iterative Proportional Fitting}
  \acro{TRS}{Truncate, Replicate, Sample}
  \acro{OTP}{OpenTripPlanner}
  \acro{GTFS}{General Transit Feed Specification}
  \acro{PLM}{Pretrained Language Model}
  \acro{ABM}{Agent-based modeling}
  \acro{IPU}{Iterative Proportional Updating}
  \acro{MCMC}{Markov Chain Monte Carlo}
  \acro{LuST}{Luxembourg SUMO Traffic}
  \acro{FCD}{Floating Car Data}
  \acro{VALFRAM}{Validation Framework for Activity-based Models}
  \acro{RMSE}{Root Mean Squared Error}
  \acro{API}{Application Programming Interface}
  \acro{MATSim}{Multi-Agent Transport Simulation}
  \acro{MIT}{Motorized Individual Transport}
  \acro{VR}{Virtual Reality}
  \acro{HCI}{Human-Computer Interaction}
\end{acronym}

\section{Introduction}


Today’s society is more mobile than ever. Whether on foot, by bicycle, car, bus, or train, mobility shapes our lives. With 55\% of the global population living in cities~\cite{united_nations_department_of_economic_and_social_affairs_population_division_world_2019}, urban mobility faces critical challenges like congestion and emissions~\cite{yamazaki_association_2005, martinez-bravo_trade-offs_2019}. To address these, researchers and policymakers are exploring technology- and policy-driven innovations~\cite{ceder_urban_2021}.
Technological approaches include automated vehicles, traffic flow optimization, or new modes of transportation such as e-scooters or urban air mobility~\cite{ceder_urban_2021}. 
Policy is another driver of more sustainable, accessible, and equitable urban mobility. Examples include fare-free public transport~\cite{cats_prospects_2017}, or serious investments in cycling infrastructure like recently decided on in Paris~\cite{hidalgo_plan_2024}.
%
A fundamental challenge lies in predicting which innovations will succeed.
This is particularly complex for individual mobility choices, where human factors are crucial~\cite{sadeghian_designing_2022}. Decisions are shaped by various factors, including needs and preferences, as well as sociodemographic and socioeconomic factors. This makes it challenging to predict personal mobility choices. 
For example, ride-hailing was expected to reduce private car ownership and ease traffic congestion by turning idle vehicles into shared resources. Empirical studies, however, report little change in car ownership, but instead increases in congestion and delays, and significantly decreased public transportation usage~\cite{diao_impacts_2021}.

To better anticipate the consequences of new mobility innovations, decision makers often rely on empirical studies and traffic simulations before any real-world deployment. These methods can be effective, but they are typically applied only after the design space has already been narrowed to a few concrete alternatives.
Today, upfront user studies and subsequent traffic simulations remain the standard to understand personal mobility decisions and overall traffic behavior~\cite{bigi_evaluating_2023, krajzewicz_embedding_2018}.
The simulations are built on complex rule sets tailored to specific scenarios and assumptions derived from the user studies. 
When carefully designed, these studies provide valuable prior knowledge of individual mobility choices and offer a solid foundation for simulation models.
However, this workflow is poorly suited for the \emph{early stages} of the design and decision-making process, where potentially disruptive ideas need to be explored. 
We use the term \textit{early-stage} to describe a phase in which researchers aim to obtain predictions for emerging concepts as precise as possible, but generated quickly and with limited resources. 
In this phase, designers and decision makers need rapid, low-cost feedback to screen ideas, identify promising directions, and refine hypotheses before committing to extensive data collection or costly full-scale modeling.
Current methods do not support this. User studies require considerable recruitment, coordination, and analysis effort, often taking months to complete~\cite{ermes_mobilitat_2020}. Consequently, high costs leave many promising innovations untested.
Furthermore, current simulation approaches~\cite{schrab_large-scale_2023, codeca_luxembourg_2017} impose structural constraints: they rely on predefined behavioral rules, fixed activity schedules, and probabilistic choice models that restrict how agents can act. 
As a result, these models struggle to capture the flexibility, contextual reasoning, and interpretability needed to understand how and why people make mobility decisions.

\ac{LLM}-based generative agents~\cite{park_generative_2023, park_social_2022, liu_toward_2024} have emerged as a method for simulating believable human behavior, producing plausible natural-language reasoning for their decisions. 
Motivated by this work, early transport-oriented studies have begun adopting such agents to support simulation setup or route choice modeling. 
However, these approaches still evaluate agents mainly on narrative plausibility; they are rarely tested on whether they can reproduce empirically validated, large-scale human behavior. 
As a result, existing traffic-related systems lack sociodemographic grounding and seldom embed agents in environments where behavior must respect real-world constraints, leading to mobility patterns that are not validated against census data or large-scale mobility surveys (see \autoref{tab:gta_comparison}).

To address these challenges, we take a step toward plausible \emph{and} empirically grounded mobility simulations using \ac{LLM}-based agents. 
With \ac{GTA}, we show that \textbf{LLM agents can reproduce real-world mobility decisions and traffic patterns when grounded in population data and embedded in a realistic urban environment}. 
We position this as a foundational contribution for HCI: reliable, interpretable simulations of human behavior are a prerequisite for using generative agents to support early-stage design work, explore novel mobility innovations and policies, or reason about future mobility-related user experiences.

To achieve this goal and unlike traditional agent-based traffic simulations that rely on fixed data distributions, survey-derived activity patterns, and probabilistic choice models, \ac{GTA} builds agent personas from census-aligned microdata and positions them into a simulation environment in which mobility behavior can only emerge under real-world constraints (i.e., the topology of the city or the availability of modalities and routes as defined by the traffic network and transit schedules).
Personas are sampled from \textbf{census-aligned microdata}. Integrated with the traffic simulation \ac{SUMO}, \ac{GTA} produces realistic traffic flows with minimal manual tuning. We ground and evaluate the system with real data, aligning personas to the "Mobility in Germany 2017"~\cite{ermes_mobilitat_2020} distributions, and benchmarking outputs against travel survey indicators and official traffic counts. This grounding moves \ac{GTA} beyond narrative plausibility toward empirically validated behavioral realism, a prerequisite for any future use of generative agents in evaluating new mobility technologies or HCI concepts.

Our evaluation shows that agents plan believable daily activity schedules.
When integrated with the traffic simulation \ac{SUMO}, they make realistic mode choices for their trips, and traffic simulations generated with \ac{GTA} resemble real traffic.  We further show that generative traffic agents are well-suited to capture the impact of sociodemographic factors on individual mobility behavior.
\ac{GTA}'s simulated mobility behavior aligns with patterns reported in large-scale, survey-based mobility studies~\cite{ermes_mobilitat_2020}. Despite promising results, our evaluations also reveal systematic deviations.

\medskip

\noindent\fcolorbox{orange}{orange!30}{\textbf{Contribution Statement}~\cite{Wobbrock.2016}:}

\smallskip

\begin{itemize}[leftmargin=*]
    \item \textbf{Artifact or System \& Methodological.} \ac{GTA}, a novel method for simulating realistic, context-sensitive mobility behavior based on persona-driven LLM agents grounded in population statistics.
    \item \textbf{Empirical Study.} We evaluate \ac{GTA} in large-scale simulations of Berlin, comparing generated behaviors against empirical mobility data to assess how well \acp{LLM} capture the influence of personal and socioeconomic factors on transportation choices.
\end{itemize}

\section{Related Work}

\begin{table*}[]
  \centering
  \footnotesize
  \caption{Key properties of \ac{GTA} compared to prior work.}
  \label{tab:gta_comparison}
  {\renewcommand{\arraystretch}{1.2}%
  \begin{tabularx}{\linewidth}{p{2cm}|p{1.5cm}p{1.8cm}p{1.2cm}p{1.2cm}p{1.2cm}p{1.2cm}p{2cm}p{1.2cm}p{1cm}}
    \toprule
    \textbf{Approach} &
    \textbf{City} &
    \textbf{Peer-reviewed (venue)} &
    \textbf{Uses \ac{LLM} agents} &
    \textbf{Persona-based agents} &
    \textbf{Grounded in mobility or user data} & \textbf{Validated against real-world data} &
    \textbf{City-scale transport model (traffic simulation)} &
    \textbf{Supports policy or design evaluation} &
    \textbf{Open source} \\
    \midrule
    Free Public Transport ABM~\cite{bigi_evaluating_2023} &
    Luxembourg &
    \cmark (TRR)& 
    \xmark &
    \xmark &
    \cmark &
    \cmark  &
    \cmark~(MATSim) &
    \cmark &
    \xmark \\
    Intermodal ABM~\cite{krajzewicz_embedding_2018} &
    n/a &
    \cmark (Procedia)& 
    \xmark &
    \xmark &
    \cmark &
    &
    \cmark~(TAPAS + SUMO) &
    \cmark &
    \xmark \\
    Social Simulacra~\cite{park_social_2022} &
    Fictional &
    \cmark (UIST)& 
    \cmark &
    \cmark &
    \xmark &
    \xmark &
    \xmark &
    \xmark &
    \xmark \\
    Generative Agents~\cite{park_generative_2023} &
    Fictional &
    \cmark (UIST)& 
    \cmark &
    \cmark &
    \xmark &
    \xmark &
    \xmark &
    \xmark &
    \cmark - \href{https://github.com/joonspk-research/generative_agents}{Link} \\
    LLM Mode Choice~\cite{liu2024travel_behavior} &
    n/a &
    \xmark & 
    \cmark &
    \xmark &
    \cmark &
    \xmark &
    \xmark &
    \cmark &
    \xmark \\
    LingoTrip~\cite{qin2025lingotrip} &
    Hong Kong &
    \cmark (J. Public Trans)& 
    \cmark &
    \xmark &
    \cmark &
    \xmark &
    \xmark &
    \cmark &
    \xmark \\
    Agentic Route Choice LLM~\cite{wang2025agentic_routes} &
    n/a &
    \cmark (TR Part C)& 
    \cmark &
    \cmark &
    \cmark &
    \xmark &
    \xmark &
    \cmark &
    \cmark - \href{https://github.com/georgewanglz2019/LLMTraveler}{Link} \\
    Synthetic \ac{HCI} Data~\cite{hamalainen2023evaluating} &
    n/a &
    \cmark (CHI) & 
    \cmark &
    \xmark &
    \cmark &
    \xmark &
    \xmark &
    \cmark &
    \xmark \\
    Synthetic Personae in \ac{HCI}~\cite{prpa2024challenges} &
    n/a &
    \cmark (CSCW)& 
    \cmark &
    \cmark &
    \cmark &
    \xmark &
    \xmark &
    \cmark &
    \xmark \\
    Persona Effect~\cite{hu2024persona_effect} &
    n/a &
    \cmark (ACL)& 
    \cmark &
    \cmark &
    \cmark &
    \xmark &
    \xmark &
    \cmark &
    \cmark - \href{https://github.com/cambridgeltl/persona_effect}{Link} \\
    ChatSUMO~\cite{li2024chatsumo} &
    Albany &
    \cmark (IEEE TIV)& 
    \cmark &
    \xmark &
    \cmark &
    &
    \cmark~(SUMO) &
    \cmark &
    \xmark \\
    Cognitive Agents in Urban Mobility~\cite{calderon2025cognitiveAgents} &
    Valencia &
    \cmark (Sensors)& 
    \cmark &
    \cmark &
    \cmark &
    \xmark &
    \cmark~(SimFleet) &
    \cmark &
    \cmark - \href{https://github.com/javipalanca/simfleet}{Link} \\
    GATSim~\cite{liu2025gatsim} &
    Test Network &
    \xmark & 
    \cmark &
    \cmark &
    \xmark &
    \xmark &
    \xmark~(custom) &
    \cmark &
    \cmark - \href{https://github.com/qiliuchn/gatsim}{Link} \\
    LLM-based Synthetic Urban Residents~\cite{wang2024largelanguagemodelsurban} &
    Tokyo &
    \cmark (NeurIPS)& 
    \cmark &
    \cmark &
    \cmark &
    \xmark &
    \xmark &
    \cmark &
    \cmark  - \href{https://github.com/Wangjw6/LLMob/}{Link} \\
    LLM-agent-based Framework~\cite{liu2024llmagentbasedmodelingtransportationsystems} &
    Miniature Metro Area &
    \xmark & 
    \cmark &
    \cmark &
    \xmark &
    \xmark &
    \cmark~(DTALite) &
    \cmark &
    \xmark \\ \hdashline
    \textbf{\ac{GTA} (This Work)} &
    \textbf{Berlin} &
    \cmark (CHI) &
    \cmark &
    \cmark &
    \cmark &
    \cmark &
    \cmark~\textbf{(SUMO)} &
    \cmark &
    \cmark \textbf{- \href{https://github.com/ciao-group/Generative-Traffic-Agents}{Link}}\\
    \bottomrule
  \end{tabularx}
  }%
\end{table*}

Technical mobility innovations in HCI~\cite{european_commission_sustainable_2020, shaheen_mobility_2016, standing_implications_2019} and policy interventions such as pricing, fares, and infrastructure changes~\cite{hidalgo_plan_2024, cats_prospects_2017, haarstad_beyond_2022} depend on robust, empirically grounded simulations to judge effects before deployment. We provide such simulations with \ac{GTA} by linking survey-grounded personas to city-scale traffic models and LLM-based generative agents.

\subsection{Traffic Simulations and Scenarios}

Traffic simulations are commonly categorized by their resolution. Macroscopic models describe flows, densities, and speeds on network links, while microscopic simulators track individual vehicles and interactions (e.g., lane-changing~\cite{alghamdi_comparative_2022}). Mesoscopic approaches occupy the middle ground. Agent-based models represent individual travelers with activity schedules and mode choices.
For Berlin, \citet{ziemke_matsim_2019} used the mesoscopic simulator MATSim~\cite{matsim_reference_2016} to build a synthetic population and activity-travel patterns from census and survey data. \citet{schrab_large-scale_2023} transferred this scenario to \ac{SUMO}, yielding a microscopic network with 2.2\,million detailed trips. Such scenarios are powerful but difficult to modify: most behavioral variation is encoded via fixed utility functions and distributions. Adding new technologies or policies often require new estimation runs, new surveys, or substantial engineering effort~\cite{nguyen_overview_2021}. Interdisciplinary researchers wanting to explore “what-if” scenarios face steep learning curves and limited access to calibration workflows.
Recent work combines traffic simulation with \ac{LLM}-enabled systems to lower these barriers. ChatSUMO integrates an \ac{LLM} with \ac{SUMO} to generate traffic scenarios from natural language, import road networks from OpenStreetMap, and run simulations without requiring expert knowledge of configuration files~\cite{li2024chatsumo}. Other work enriches multi-agent traffic models with cognitive, \ac{LLM}-driven agents that exhibit memory and adaptation under transport disruptions~\cite{calderon2025cognitiveAgents}. These approaches reduce technical friction and illustrate how generative tools can accelerate simulation setup, but still rely on conventional demand models and do not systematically validate agent behavior against city-specific mobility data. \ac{GTA} builds directly on existing Berlin scenarios, yet replaces fixed decision rules with \ac{LLM} personas whose planning and mode choice behavior are calibrated against real-world mobility statistics.

\subsection{LLM-Based Generative Agents}

\acp{LLM} show strong performance in text generation and in-context reasoning across a wide range of tasks~\cite{openai_gpt-4_2024,dubey_llama_2024,zhao_survey_2023,wang_survey_2024}. Early work on “generative agents” demonstrated that LLM-driven characters can form memories, pursue goals, and interact credibly in simulated social environments~\cite{park_social_2022, park_generative_2023}. Surveys on \ac{LLM}-based agents and agent-based simulation argue that LLM agents can address classic challenges of agent-based modeling, including perception, heterogeneity, and adaptive decision-making~\cite{gao_large_2023}. These works establish that \ac{LLM} agents can support open-ended scenario exploration but provide limited evidence on how well they reproduce measured human behavior.

In transport research, several works examine LLMs as models of travel behavior. Liu et al.\ investigate whether LLMs can reproduce observed mode choice patterns and show that raw, zero-shot \ac{LLM} responses misalign with human data but that few-shot calibration and persona prompts can improve alignment~\cite{liu2024travel_behavior}. Qin et al.\ propose LingoTrip, a spatiotemporal prompting scheme for predicting individual next trips with an \ac{LLM} and report advantages over state-of-the-art prediction models in terms of accuracy and data efficiency~\cite{qin2025lingotrip}. More recently, Wang et al.\ introduce an “agentic” \ac{LLM} framework for day-to-day route choice modeling, where \ac{LLM} agents iteratively adjust their choices in a dynamic network setting~\cite{wang2025agentic_routes}. Together, these studies treat \acp{LLM} as behavioral models for specific choice problems, but they do not embed agents into a full city-scale traffic simulation that is directly comparable to a baseline scenario.

HCI has also developed \ac{LLM}-based synthetic participants and personas. Hämäläinen et al.\ show that \acp{LLM} can generate synthetic HCI study data that resemble human responses, but warn that such data are better suited for early-stage ideation than for replacing participants~\cite{hamalainen2023evaluating}. Prpa et al.\ discuss opportunities and risks of \ac{LLM}-based synthetic personae and data in HCI, including issues of bias, validity, and over-reliance on synthetic samples~\cite{prpa2024challenges}. Hu and Collier analyze the “persona effect” and quantify how persona prompts change \ac{LLM} outputs, but also demonstrate that these shifts only partially explain human variation~\cite{hu2024persona_effect}. Across this literature, synthetic agents are framed as complements to, not substitutes for, human data. Unlike prior generative-agent transport systems such as GATSim~\cite{liu2025gatsim}, which focus on cognitively rich agents in stylized networks, GTA targets city-scale scenarios and explicitly calibrates LLM-driven personas against census and survey data in Berlin.

\ac{GTA} connects these strands. Conceptually, it follows generative-agent work by using \ac{LLM}-driven personas with goals, memories, and constraints. Methodologically, it is closer to transport studies that calibrate \ac{LLM} behavior against measured mode choice and trip patterns. Unlike prior HCI work on synthetic participants, \ac{GTA} agents are not only plausible but also explicitly anchored in census statistics and a Berlin mobility survey; their daily schedules and mode choices are validated against empirical distributions. Unlike prior \ac{LLM}–traffic integrations, our agents operate inside a large-scale \ac{SUMO} scenario and can be evaluated at both the individual (persona) and aggregate (traffic) levels. This dual focus enables us to examine when persona-based generative agents are useful for mobility research and where they deviate in systematic, interpretable ways.

\section{Introducing GTA: Generative Traffic Agents}

\begin{figure*}[]
    \centering
        \includegraphics[width=0.5\linewidth]{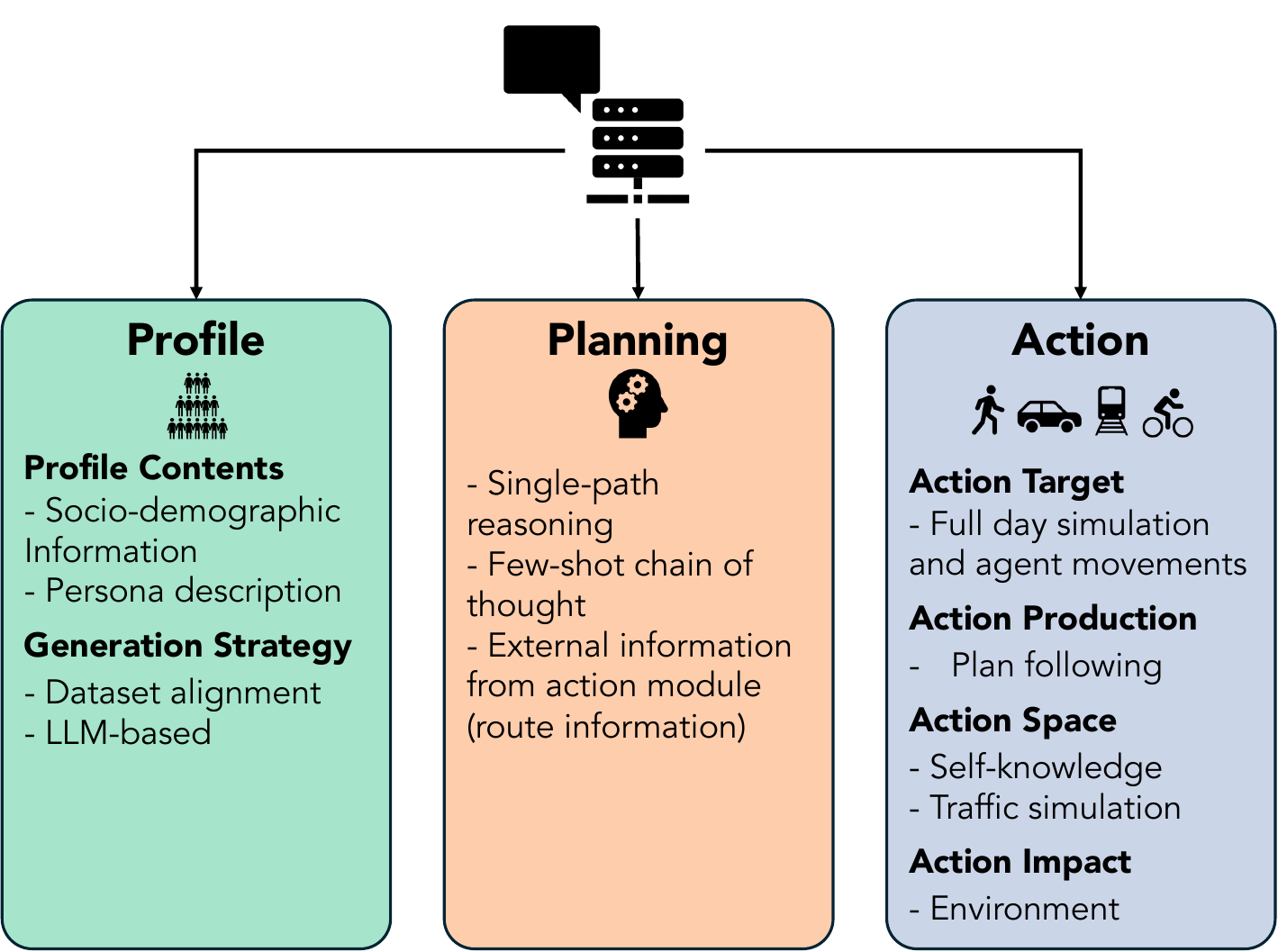}
        \caption{GTA's modularized architecture.}
        \label{fig:architecture_modules}
        \Description{A three-part diagram showing the modular structure of generative agents, divided into Profile, Planning, and Action components.}
\end{figure*}

The design of \ac{GTA} is guided by three goals that enable rapid, early-stage, yet empirically grounded mobility simulation.
First, \textbf{population-grounded diversity}: agents should reflect sociodemographic variation without relying on handcrafted rules derived from costly user studies. We therefore combine census-aligned microdata with persona-driven \acp{LLM} to generate heterogeneous preferences and day plans.  
Second, \textbf{contextual decision-making under real-world constraints}: agents must reason about
actual distances, travel times, and available modes. Integrating with \ac{SUMO} (and \ac{OTP}) enables
mode choice and routing based on concrete, network-informed options rather than abstract prompts. 
Third, \textbf{interoperability}: to support HCI and mobility
researchers, \ac{GTA} is designed as a modular pipeline that operates on \ac{OSM}/\ac{GTFS} data and
produces outputs comparable to standard validation datasets (e.g., modal split, trip lengths,
traffic counts). We decided to use \ac{SUMO} as it is frequently used in HCI~\cite{colley2024pedsumo, colley2025uam} and allows GTA to be extended easily.

\begin{figure*}[]
        \includegraphics[width=\linewidth]{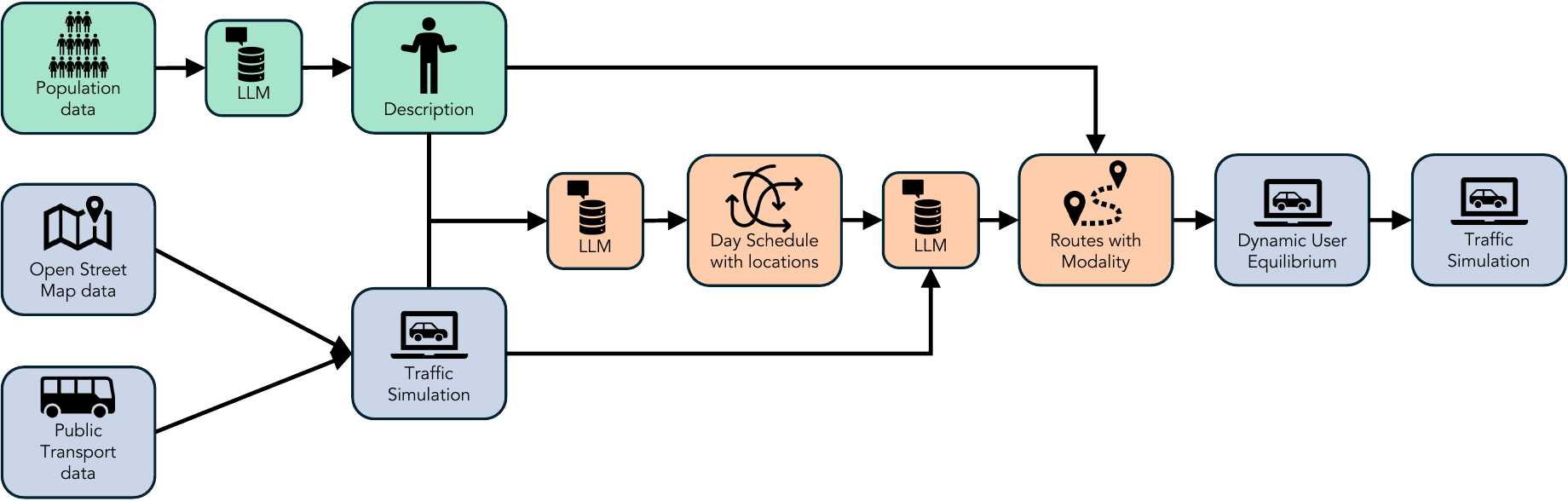}
        \caption{Information flow diagram between model components.}
        \label{fig:architecture_flow}
        \Description{A flow diagram that illustrates how different data sources and model components interact in the generative traffic simulation pipeline. It features three types of data inputs—population data, map data, and public transport data—and follows the flow of information through LLM-based modules toward traffic simulation.}
\end{figure*}

These goals motivate \ac{GTA}'s three core components (see \autoref{fig:architecture_modules}). The \textit{profile module} samples
socio-demographic attributes from census-calibrated microdata and generates matching textual
persona using an \ac{LLM}. The \textit{planning module} produces full-day activity schedules,
assigns concrete destinations based on \ac{OSM} categories, and selects transport modes using
persona information and network-informed route options. The \textit{action module} instantiates
these plans by computing multimodal routes, refining them through the dynamic user equilibrium in
\ac{SUMO}, and executing the final traffic simulation on \ac{OSM}/\ac{GTFS}-derived networks. \autoref{fig:architecture_flow} shows how the model components are connected.

\subsection{Profile Module: Population Sampling and Agent Generation}\label{ch:profile}

Each agent receives a profile consisting of socio-demographic attributes and a natural-language persona description. Attributes are first sampled from population statistics, after which an \ac{LLM} generates a matching persona description.

\paragraph{Population Sampling}

The synthetic population is based on anonymized nationwide census data~\cite{ermes_mobilitat_2020}. Each survey record includes a calibrated weight reflecting its representation in the population, adjusted through design weights, non-response correction, and post-stratification to known margins~\cite{mid2017handbuch,mid2017methodik}.

To generate a synthetic population of size \(N\), we scale the weights \(w_i\) such that \(\sum_i w_i^* = N\), and apply \ac{TRS}~\cite{LOVELACE20131} to convert the fractional weights into integer replication counts. TRS handles the trade-off between accuracy and computational cost.
While we focus on the "Mobility in Germany 2017" dataset due to its regional relevance and detail, our approach generalizes to any census data. If only marginal distributions are available, joint distributions can be inferred using statistical techniques like \ac{IPF}~\cite{deming1940least}.

\paragraph{Description Generation}

Each synthetic agent's attributes (e.g., household income) are provided in the \ac{LLM} prompt generating the persona description (see Listing~\ref{lst:description_prompt}). These personas enrich the simulation with personality cues while preserving demographic realism.

\subsection{Planning Module: Activity Scheduling and Trip Planning}\label{ch:Planning}

To generate realistic travel demand, agents construct individual full-day activity plans prior to traffic simulation. This includes a sequence of activities with assigned times and locations, plus transport modality for each trip. These activities are generated based on the persona description and sociodemographic attributes. Pre-planning reduces runtime and \ac{LLM} queries, and enables diverse, yet tractable, behavioral modeling.
While trip execution is handled by SUMO, activity planning depends on mobility constraints such as location availability and travel durations. Consequently, the boundary between planning and action is not as strict as \autoref{fig:architecture_modules} suggests.

\paragraph{Activity Scheduling}

Each agent’s schedule specifies the activity type, timing, and associated location. To ensure spatial plausibility, location candidates are extracted from \ac{OSM}, using attributes such as \texttt{building}, \texttt{amenity}, and \texttt{shop}. These are grouped into predefined location categories appropriate for different activity types. Activities are only linked to locations within these categories, selected according to the sampling strategy described in \autoref{ch:actions:location_choice}.

Temporal and contextual variation is captured by embedding structured prompts with factors like weekday/weekend distinctions~\cite{raux2016variability,dobler2021patterns,schlich2003habitual}, policy interventions~\cite{cats_prospects_2017}, or major events~\cite{giuliano2021analyzing}. Based on the agent’s persona, date, and valid \ac{OSM} location categories, the \ac{LLM} returns a structured activity schedule. Agents are instructed to avoid mentioning transportation or commuting behaviors in the plan to maintain a clean separation between abstract planning and concrete mobility execution.

\paragraph{Trip Planning}  
After generating daily schedules, we resolve activities requiring travel by mapping their types to predefined building categories and sampling concrete locations (see \autoref{ch:actions:location_choice}).
\ac{SUMO} estimates travel times and distances for all modes of transportation to compute feasible travel options. We provide the \ac{LLM} with the agent's persona, identified trip segments, and candidate routes. Using few-shot \ac{CoT}, the model is given domain knowledge and example reasoning steps, enabling it to evaluate the available route options and select the mode and route that best match the agent’s persona and the temporal and spatial characteristics of each candidate. See the prompt in \autoref{lst:transport_prompt} in \autoref{ch:appendix}.

\subsection{Action Module: Traffic Simulation and Dynamic Routing}\label{ch:action}

The action module supports planning by (1) defining building categories for each activity type, (2) selecting specific destination buildings based on agent context, and (3) computing multi-modal routes between them. It also executes the simulation and performs dynamic user equilibrium routing~\cite{sumo_reference_paper_2018}, ensuring agents adapt their routes to congestion rather than following static shortest paths.

\paragraph{Location Choice}\label{ch:actions:location_choice}

Activity destinations (e.g., for work or shopping) are selected using a proximity-based rule: agents choose the nearest building of a suitable type. While simplistic, this approach best matched empirical trip length distributions in early experiments. More complex methods (e.g., distance-decay weighting) shifted the distribution toward longer trips, worsening alignment with real-world data. Although trips $<$\SI{1}{\kilo\metre} are still underrepresented, the nearest-only rule offers a good trade-off between computational efficiency and realistic destination choice.
Relevant building types are defined via \ac{OSM} tags---including \texttt{building}, \texttt{amenity}, \texttt{office}, \texttt{shop}, and \texttt{craft}---and mapped to categories like workplaces or leisure venues (see \autoref{lst:exemplary_building_cat}). During trip plan generation, the \ac{LLM} selects activity types based on these categories.
Home locations are sampled using an area-weighted heuristic: each residential building receives a probability $p_i$ proportional to its floor area, assuming larger buildings house more people: $p_i = \frac{\text{area}_i}{\sum_j \text{area}_j}$.

\paragraph{Initial Routing}

To inform mode choice, initial route options are computed per trip and modality. For car, bike, and walk trips, \ac{SUMO} estimates shortest paths. For public transport, \ac{OTP} computes multimodal, time-dependent transit routes from \ac{OSM} and \ac{GTFS} data. Both methods ignore traffic congestion and serve as approximations; final routing is refined later through dynamic equilibrium.

\paragraph{Scaling}

Full-scale simulation of Berlin’s population ($\approx3.9$ million inhabitants) is computationally very expensive. With 8 CPUs and 8 Nvidia A30 GPUs, simulating 10\% of the population takes \SI{43}{\hour}, while equilibrium computation can take weeks~\cite{schrab_large-scale_2023}. We thus simulate 10\% of the population and scale agents by weight, resampling locations within a \SI{2}{\kilo\metre} radius to preserve spatial diversity.

\paragraph{Dynamic User Equilibrium and Final Simulation}

Final route assignments are computed by iteratively adjusting paths to approximate dynamic user equilibrium~\cite{sumo_reference_paper_2018} to capture network effects and congestion, yielding more realistic travel patterns in the final simulation.

\subsection{Simulation Setup}\label{methods:sim_env}

To enable agents to reason about the traffic environment, such as the durations and lengths of potential routes, and to simulate the resulting traffic flow, \ac{GTA} integrates external traffic simulators and trip planners.
\ac{SUMO}~\cite{sumo_reference_paper_2018} supports various modes of transportation, including walking, cycling, driving, and public transit, and offers high-resolution, vehicle-level simulation. This granularity is particularly valuable for analyzing the impact of disruptions, policy interventions, and mobility innovations, as it captures detailed effects such as speed adjustments, rerouting, and congestion dynamics.
GTA’s modular architecture also allows alternative services for route planning (e.g., \ac{OTP}~\cite{OpenTripPlanner2}). 

\subsection{Urban Mobility Behavior - Case Study Berlin}

To evaluate the impact of mobility innovations, simulations must capture how people move within cities, which transport systems they use, and how sociodemographic factors shape their choices. We use Berlin as case study because it shares key challenges with other global cities, such as socioeconomic pressures and inner-city gentrification~\cite{holzl_vulnerable_2021,huning_social_2015}, while offering rich census and mobility data for validation~\cite{ermes_mobilitat_2020}. Like Shanghai~\cite{magdolen_combining_2019} or London~\cite{rode_towards_2015}, Berlin faces increasing demand for sustainable and equitable mobility solutions.  

86\% of Berlin’s 3.9 million residents are mobile daily, making an average of 3.1 trips that cover 39 kilometers and take 97 minutes~\cite{ermes_mobilitat_2020}. The modal split reflects a diverse mobility landscape: 34\% car, 27\% walking, 25\% public transport, and 15\% cycling. These patterns are strongly shaped by socioeconomic status. Higher-income groups rely more on cars and less on public transport; students (40\%) and unemployed individuals (37\%) are the heaviest public transport users. Older adults cycle less and shift away from transit. These trends highlight the heterogeneity of mobility choices, underscoring the need for simulations that account for demographic diversity.

\section{Evaluating GTA}
We evaluate \ac{GTA} by comparing simulation outputs against traffic counts and survey data~\cite{ermes_mobilitat_2020}.

\subsection{Experimental Setup}

To evaluate \ac{GTA}, we ran experiments that vary in scope, model components, and simulation configurations. We compare the results against empirical data to assess plausibility.

\paragraph{Simulation Scope and Population Sampling}
For efficiency, we simulate a 1\% sample of Berlin’s population (35,769 agents) without scaling. While we generate personas, activity schedules, trip plans, and simulate mode choices for this sample, we do not run the final traffic simulation for Berlin due to the substantial manual corrections required to prepare the \ac{SUMO} network~\cite{schrab_large-scale_2023}.
Instead, we run the full simulation in the Wedding district using a 10\% sample (8,680 agents), where the smaller network allowed the required manual corrections. For realistic traffic counts, we scale agents to represent the full 2024 population of Wedding (86,796 residents). We evaluate simulation realism by comparing modeled traffic volumes with official traffic counts, sampling at the same monitoring locations.

\paragraph{Model Variants and Ablation Studies}
By default, agents use the Llama 3.1 8B language model~\cite{grattafiori2024llama3herdmodels} and all reasoning and prompting components enabled. In one experiment, we replace it with Qwen2.5-7B~\cite{qwen2.5} to test sensitivity to model choice. All experiments are conducted with 8 NVIDIA A30 GPUs and 8 CPU cores. We evaluate three variants:
\begin{itemize}
    \item[\textbf{1a.}] \textbf{Default (Berlin, \ac{SUMO}):} Agents are modeled for the Berlin population, but the traffic simulation is not run.
    \item[\textbf{1b.}] \textbf{Default (Berlin, \ac{OTP}):} Agents are modeled for the Berlin population, but \ac{OTP} is used to generate routes.
    \item[\textbf{1c.}] \textbf{Default (Wedding, \ac{SUMO}):} Both, agent-based modeling and traffic simulations, are limited to Wedding.
\end{itemize}

To isolate the contribution of specific \ac{GTA} components, we ran these ablations:

\begin{itemize}
    \item[\textbf{2.}] \textbf{Qwen2.5-7B:} We switch to Qwen2.5-7B to test the results' dependency on the language model.
    \item[\textbf{3.}] \textbf{No system prompt:} The system prompt is removed to test explicit behavior instructions on agent reasoning.
    \item[\textbf{4.}] \textbf{Few-shot prompting without Chain-of-Thought:} Agents are presented with few-shot examples of travel behavior outputs without explicit reasoning steps to isolate the contribution of output format scaffolding.
    \item[\textbf{5.}] \textbf{Zero-shot Chain-of-Thought:} Agents are given a zero-shot prompt that encourages step-by-step reasoning without examples, testing the stand-alone effectiveness of reasoning scaffolds.
\end{itemize}

\subsection{Evaluation Methodology}

To assess the quality of our simulation outputs, we compare key indicators produced by \ac{GTA} against real-world network-level counts and individual travel behavior patterns.
For the network-level evaluation, we use traffic counts available in the Berlin Open Data Portal~\cite{berlin_traffic_counts}. To analyze travel behavior, we compare the following key indicators from the "Mobility in Germany 2017" survey~\cite{ermes_mobilitat_2020} with our results: \textit{modal split, modal split by economic status, trip length by mode, trip duration by mode.}
We compute \ac{RMSE} and an aggregate average \ac{RMSE} over modal split, trip length, and trip duration.

\paragraph{Interregional Comparison}
Lacking a direct baseline, we benchmark \ac{GTA} against mobility data from other German states. If \ac{GTA}’s error is smaller than natural interregional differences, it suggests the model captures city-specific mobility with realistic fidelity.

\paragraph{Traffic Counts}
We analyze traffic flow plausibility via hourly vehicle counts at four representative traffic count stations (see \autoref{fig:wedding_stations}) in Wedding. We normalize traffic counts to compare the shape of the daily profiles.

\section{Results}
In the following, we quantitatively evaluate \ac{GTA} against empirical data from mobility studies and traffic count data. Exemplary agent descriptions and day plans are given in \autoref{ch:appendix}.

\begin{figure}[]
	\centering
	\includegraphics[width=\linewidth]{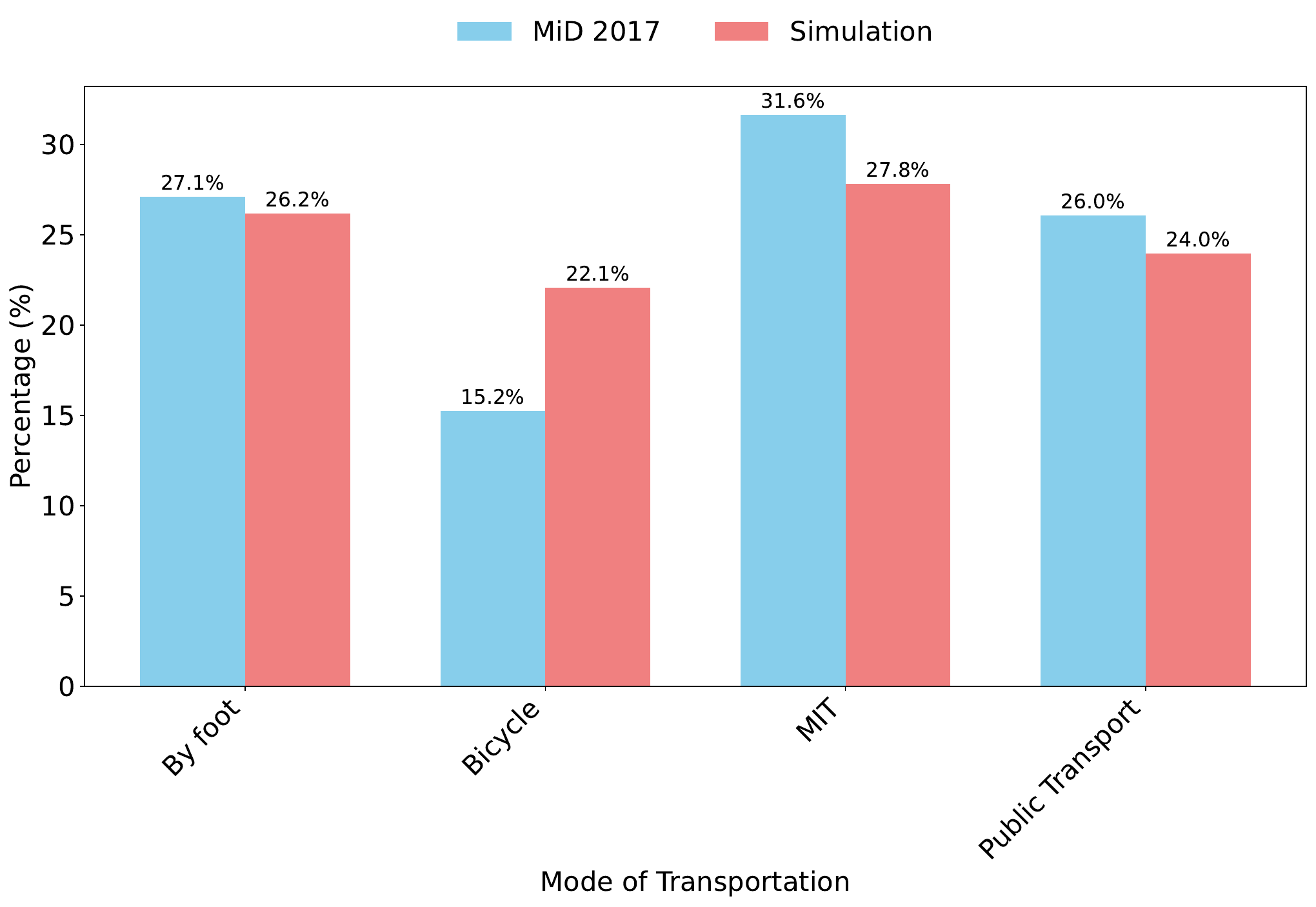}
	\caption[Modal Split.]{Modal split recorded in the survey~\cite{ermes_mobilitat_2020} and the simulation.}
	\label{fig:1a_default:modal_split}
    \Description{Grouped bar chart comparing transportation mode shares between MiD 2017 data and simulated results for walking, biking, MIT (motorized individual transport), and public transport. The simulation overestimates bicycle choice but is close for the other modalities.}
\end{figure}

\subsection{Modal Split - \autoref{fig:1a_default:modal_split}} 

\ac{GTA} closely replicates the survey's modal split~\cite{ermes_mobilitat_2020}. For instance, the share of walking trips differs by just 0.9\%, and public transport use by 2.0\%. Bicycle use is slightly overestimated, while \ac{MIT} is underestimated, with the largest deviation remaining below 7\%. Overall, the \ac{RMSE} is only 4.07, suggesting that the simulation effectively captures the distribution of mobility modes (see \autoref{fig:1a_default:modal_split}).

\begin{figure*}[ht]
    \centering
    \begin{subfigure}[t]{0.49\textwidth}
        \vspace{0pt} 
        \includegraphics[width=\linewidth]{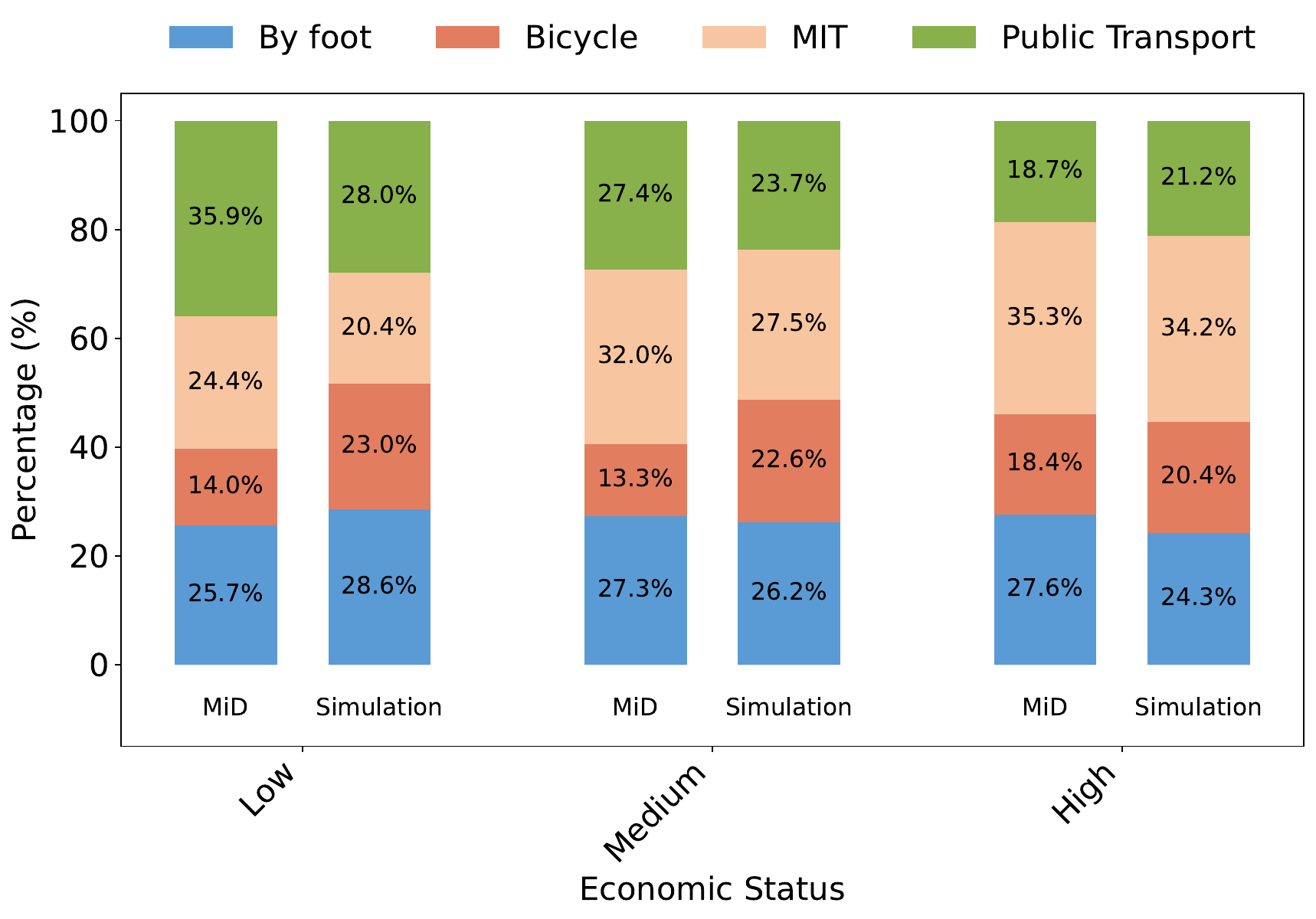}
        \caption{By \textbf{economic status}.}
        \label{fig:1a_default:modal_split_economic_status}
    \end{subfigure}
    \hfill
    \begin{subfigure}[t]{0.49\textwidth}
        \vspace{0pt} 
        \includegraphics[width=\linewidth]{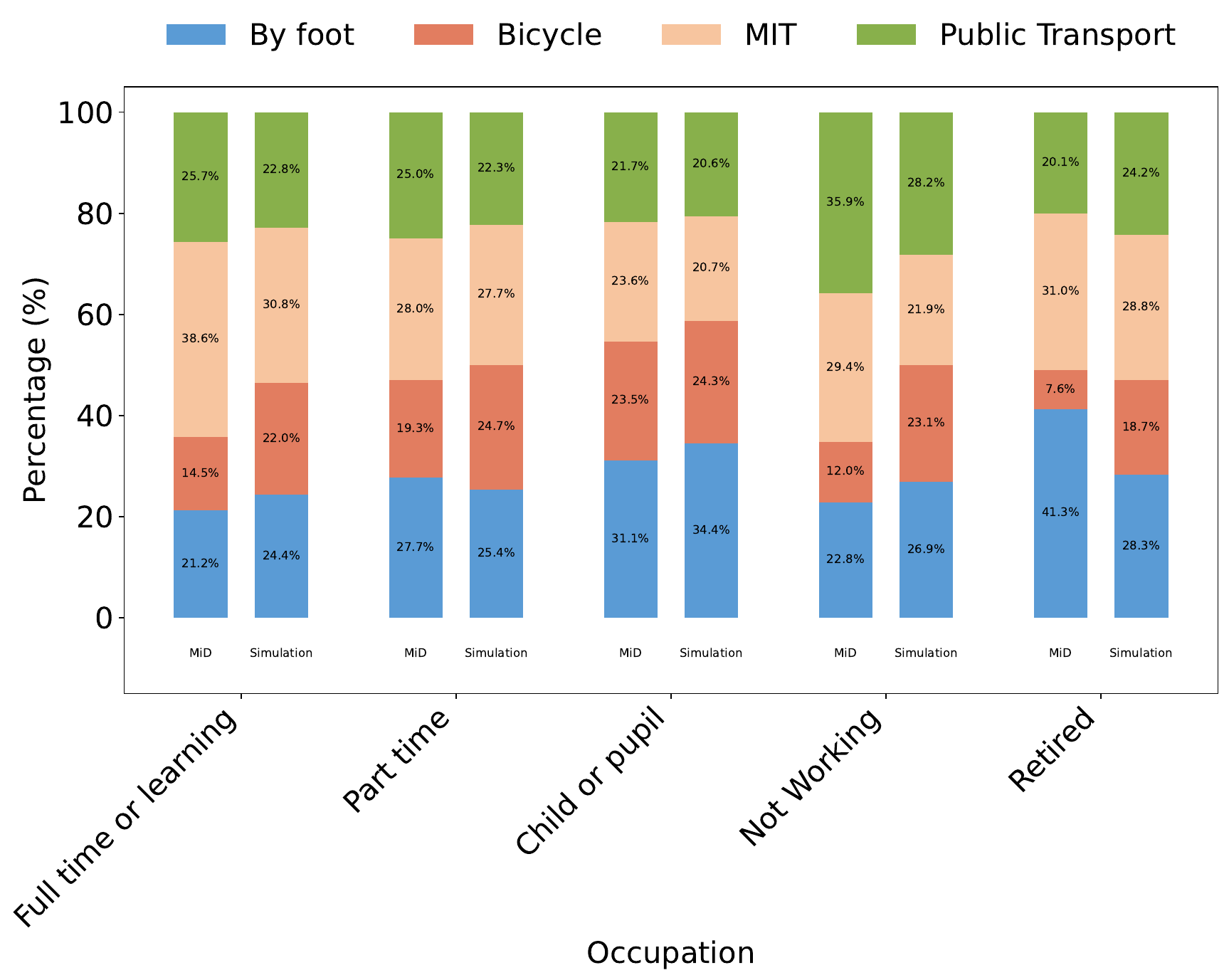}
        \caption{By \textbf{occupation}.}
        \label{fig:1a_default:modal_split_occupation}
    \end{subfigure}
    \caption{Modal split recorded in survey and simulation by (a) economic status and (b) occupation.}
    \Description{Two grouped stacked bar charts comparing transportation mode shares between MiD 2017 and simulation data by economic status (left) and occupation (right). The simulation broadly captures modal trends but differs in specific groups, often overestimating bicycle and walking use.}
\end{figure*}

\paragraph{\textbf{Modal split by economic household status} - \autoref{fig:1a_default:modal_split_economic_status}} 

The simulation reproduces several broad trends reported in the "Mobility in Germany 2017" survey~\cite{ermes_mobilitat_2020}, though deviations remain. In the simulation and the survey, public transport use declines with increasing income. Similarly, \ac{MIT} use increases with income. However, the patterns for cycling and walking differ between simulation results and survey data. While the survey shows a non-monotonic trend in bicycle use, the simulation predicts a steady decline. Walking increases slightly with income in the survey but decreases in the simulation.

\paragraph{\textbf{Modal split by occupation} - \autoref{fig:1a_default:modal_split_occupation}} 

For this comparison groups with negligible sample sizes were excluded (e.g., \textit{``employed as a part-time job or internship''} --- \emph{n} = 1 out of 3{,}206). \ac{GTA} correctly identifies the dominant modes for most groups: \ac{MIT} for full-time workers and walking for children. However, \ac{GTA} systematically overestimates active mobility. This is most visible among non-working and retired individuals: where the survey shows a strong preference for public transport and walking, \ac{GTA} shifts a significant share of these trips to cycling.

\begin{figure*}[]
    \centering
    \begin{subfigure}[b]{0.49\textwidth}
        \includegraphics[width=\linewidth]{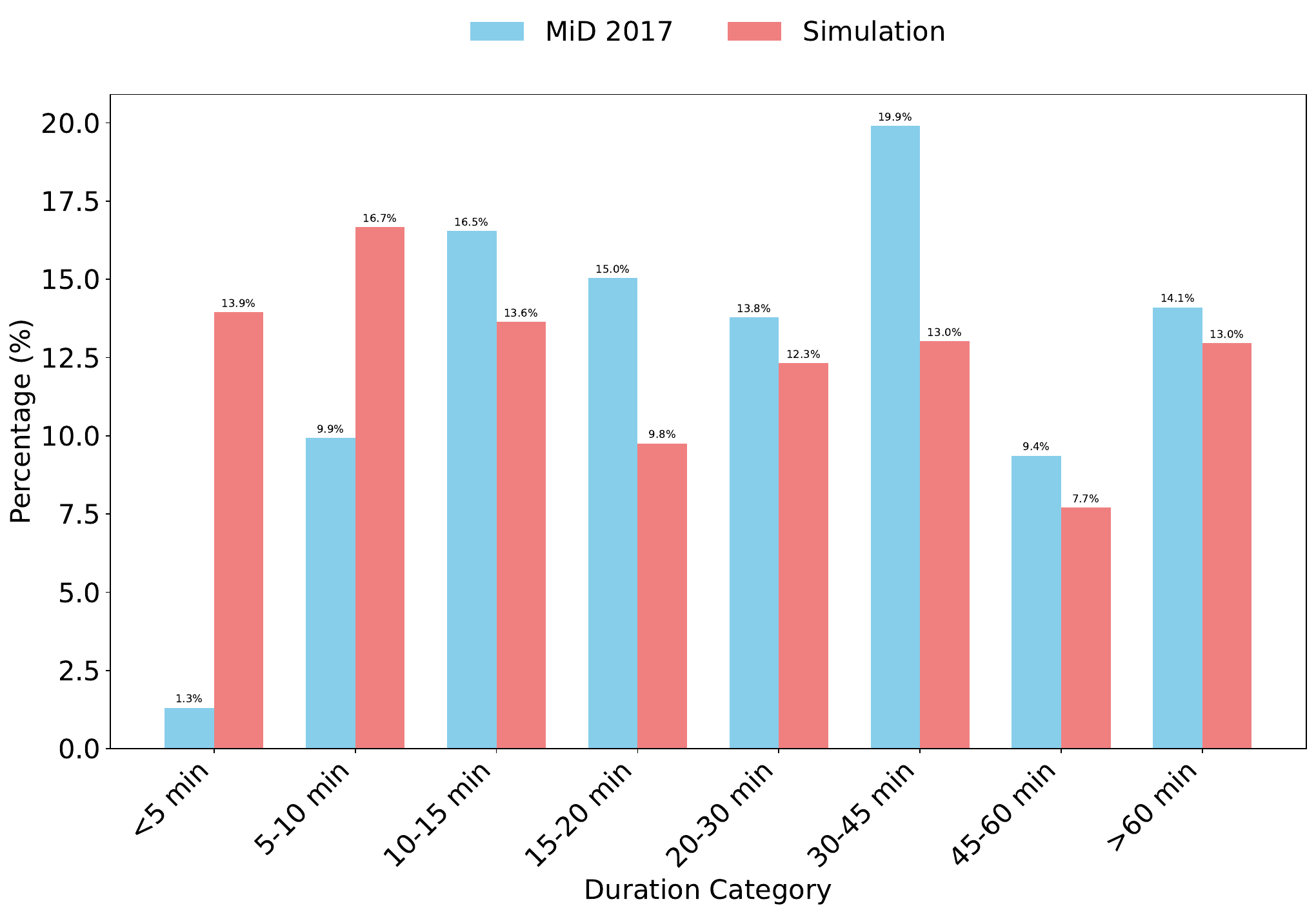}
	\caption[Route durations.]{Route durations.}
    \label{fig:1a_default:route_durations}
    \end{subfigure}
    \hfill
        \begin{subfigure}[b]{0.49\textwidth}
        \includegraphics[width=\linewidth]{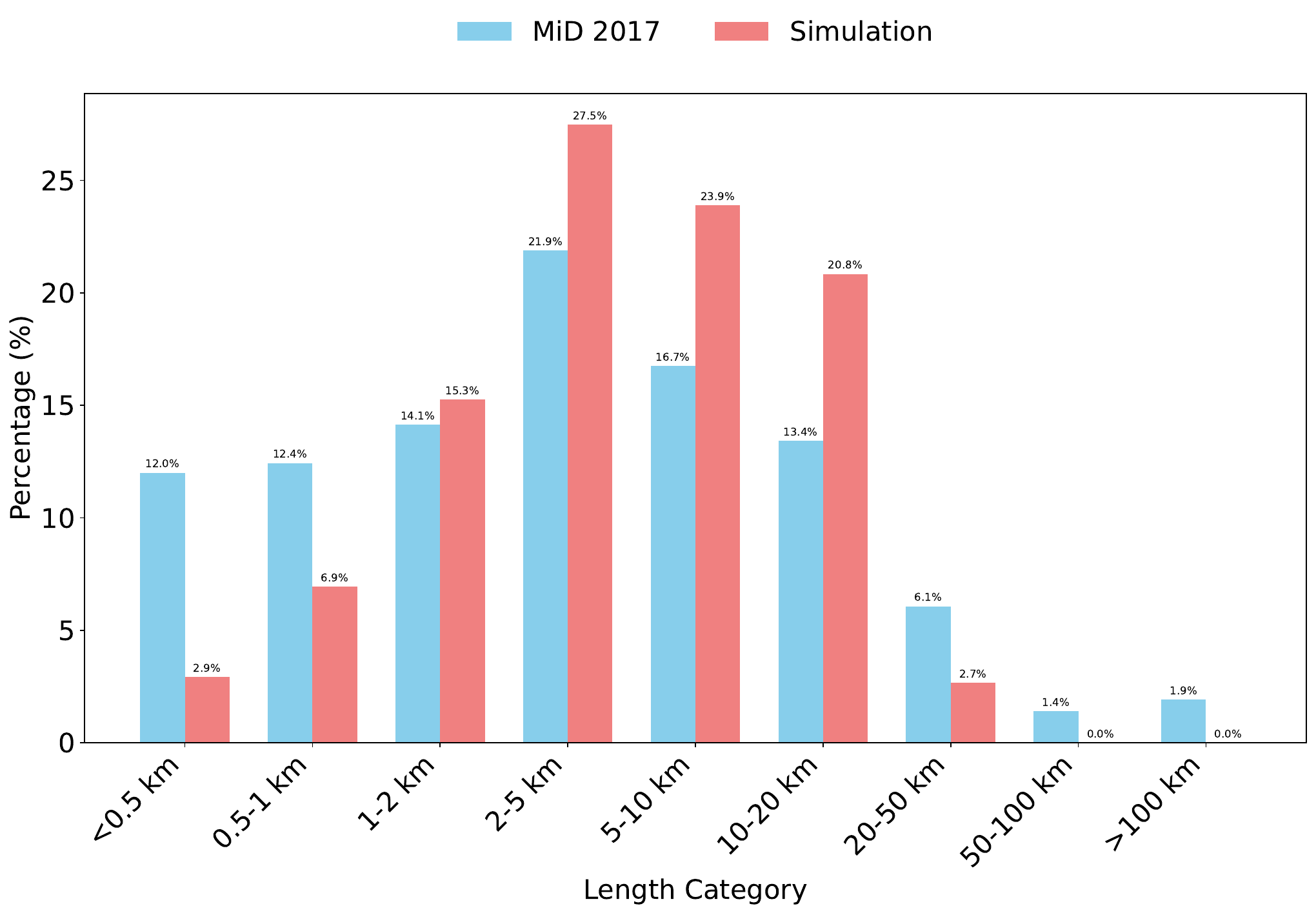}
	\caption[Route lengths.]{Route lengths.}
	\label{fig:1a_default:route_lengths}
    \end{subfigure}
    \caption{Route length and route durations for survey and simulation.}
    \label{fig:routelength}
    \Description{Two bar charts compare trip characteristics between MiD 2017 data and a simulation. The left chart shows route durations, with the simulation overestimating very short (<5 min) and 5–10 minute trips and underestimating longer trips. The right chart shows route lengths, where the simulation overrepresents trips longer than 2.5 km trips and underrepresents very short trips under 0.5 km.}
\end{figure*}

\subsection{Route Length and Duration - \autoref{fig:routelength}}

The histogram highlights a notable underrepresentation of very short trips ($<$\SI{1}{\kilo\metre}). Trips $<$\SI{0.5}{\kilo\metre} account for 2.9\% in \ac{GTA} compared to 12.0\% in the empirical data. Similarly, trips in the \SIrange{0.5}{1}{\kilo\metre} range are underrepresented (6.9\% vs. 12.4\%). \ac{GTA} overrepresents medium-range trips in the \SIrange{2}{20}{\kilo\metre} range. \SI{2}{\kilo\metre} to \SI{5}{\kilo\metre} accounts for 27.5\% of all simulated trips compared to 21.9\% in the data~\cite{ermes_mobilitat_2020}. Trips from \SI{5}{\kilo\metre} to \SI{10}{\kilo\metre} account for 23.9\% of simulated trips (vs. 16.7\%), and those from \SI{10}{\kilo\metre} to \SI{20}{\kilo\metre} make up 20.8\% (vs. 13.4\%). Very long trips are also underrepresented comprising 2.7\% of simulated trips compared to 6.1\% in the survey data~\cite{ermes_mobilitat_2020}. Trips longer than \SI{50}{\kilo\metre}  are fully absent in \ac{GTA} (0.0\% vs. 1.4\%).

\autoref{fig:1a_default:route_durations} presents the distribution of route durations. \ac{GTA} tends to produce shorter trips more frequently than in the survey, especially $<$\SI{10}{\minute}. For example, trips $<$\SI{5}{\minute} are strongly overrepresented (13.9\% vs. 1.3\%). In contrast, \ac{GTA} underrepresents trips $>$\SI{10}{\minute}. The largest gap appears for trips between \SI{30}{\minute}-\SI{45}{\minute} (13.0\% in \ac{GTA} vs 19.9\%). 

\subsection{Interregional Comparison}

\autoref{tab:state_rmse} compares travel behavior patterns across all German federal states with the simulation results for Berlin (scenario 1a). When comparing modality \ac{RMSE}, only Hamburg (1.99) has a more similar mode choice distribution to Berlin than \ac{GTA} (4.07). However, duration and length \ac{RMSE} values are consistently lower than \ac{GTA} (simulation: duration of 6.07 and length 6.12; average over states: duration of 3.58 and length 1.87). Only Bremen (4.58) and Hamburg (1.42) have lower average \ac{RMSE} values than \ac{GTA}  (5.42).

\begin{table}[]
\centering
\small
\setlength{\tabcolsep}{3pt} 
\begin{tabular}{lcccc}
\toprule
    \textbf{State} & \textbf{Modality} & \textbf{Duration} & \textbf{Length} & \textbf{Average} \\
    \midrule
 Schleswig-Holstein & 17.07 & 3.41 & 2.39 & 7.62 \\
 Hamburg & 1.99 & 1.36 & 0.91 & 1.42 \\
 Niedersachsen & 18.34 & 4.60 & 1.93 & 8.29 \\
 Bremen & 7.30 & 3.78 & 2.66 & 4.58 \\
 Nordrhein-Westfalen & 15.65 & 3.93 & 1.64 & 7.08 \\
 Hessen & 15.33 & 3.68 & 1.41 & 6.81 \\
 Rheinland-Pfalz & 18.68 & 4.08 & 2.19 & 8.31 \\
 Baden-Württemberg & 15.97 & 4.29 & 1.04 & 7.10 \\
 Bayern & 16.34 & 4.27 & 1.28 & 7.30 \\
 Saarland & 23.43 & 4.03 & 2.41 & 9.96 \\
 Brandenburg & 16.26 & 2.16 & 3.14 & 7.19 \\
 Mecklenburg-Vorpommern & 15.55 & 4.30 & 1.72 & 7.19 \\
 Sachsen & 17.12 & 3.15 & 1.00 & 7.09 \\
 Sachsen-Anhalt & 15.94 & 3.11 & 2.33 & 7.13 \\
 Thüringen & 17.62 & 3.58 & 2.03 & 7.75 \\
 \midrule
Berlin simulation (1a) & 4.07 & 6.07 & 6.12 & 5.42 \\
\midrule
Overall Mean & 15.51 & 3.58 & 1.87 & 6.99 \\
\bottomrule
\end{tabular}
\caption{RMSE between empirical data of Berlin state and other German federal states~\cite{ermes_mobilitat_2020}.}
\label{tab:state_rmse}
\end{table}

\subsection{Ablation and Variation Studies}

For experiment 1, we simulate the Wedding district and report \ac{SUMO}- and \ac{OTP}-based routing; for experiments 2-5, only \ac{SUMO} results are available. The results reveal (\autoref{tab:rmse_summary}) reveal trade-offs between improved travel-time or distance alignment and the accuracy of mode choice distributions. \ac{OTP} routing (1b) and district-level constraints (1c) improve short-trip alignment but distort overall modal splits. Similarly, Qwen2.5-7B (2) improves travel-time fit but misallocates modes. Removing \ac{CoT} scaffolding (4, 5) significantly degrades plausibility, confirming that reasoning prompts are essential for balancing mode choice accuracy.

\begin{table}[]
\centering
\small
\setlength{\tabcolsep}{3pt} 
\begin{tabular}{llcccc}
\toprule
& \textbf{Experiment} & \textbf{Modality} & \textbf{Duration} & \textbf{Length} & \textbf{Aggregated} \\
\midrule
 \textbf{1a} & \textbf{Best (default)} & \textbf{4.07} & 6.07 & 6.12 & \textbf{5.42} \\
 1b & Default (\ac{OTP}) & 8.33 & 5.63 & 4.97 & 6.31 \\
 1c & Default (Wedding) & 7.16 & 18.69 & 13.57 & 13.14 \\
 2 & Qwen2.5-7B & 11.92 & \textbf{4.44} & 8.33 & 8.23 \\
 3 & No system prompt  & 4.92 & 6.68 & 6.19 & 5.93 \\
 4 & Few-shot, no \ac{CoT} & 11.69 & 9.17 & \textbf{3.32} & 8.06 \\
 5 & Zero-shot \ac{CoT} & 7.22 & 6.93 & 5.51 & 6.55 \\
\bottomrule
\end{tabular}
\caption{RMSE of modal split, trip durations, trip lengths, and aggregated metrics for each experiment compared to the survey data~\cite{ermes_mobilitat_2020} for Berlin. The best RMSE value and the best overall experiment are bold.}\label{tab:rmse_summary}
\end{table}

\begin{figure*}[]
    \centering
    \begin{subfigure}[b]{0.47\textwidth}
\includegraphics[width=\linewidth]{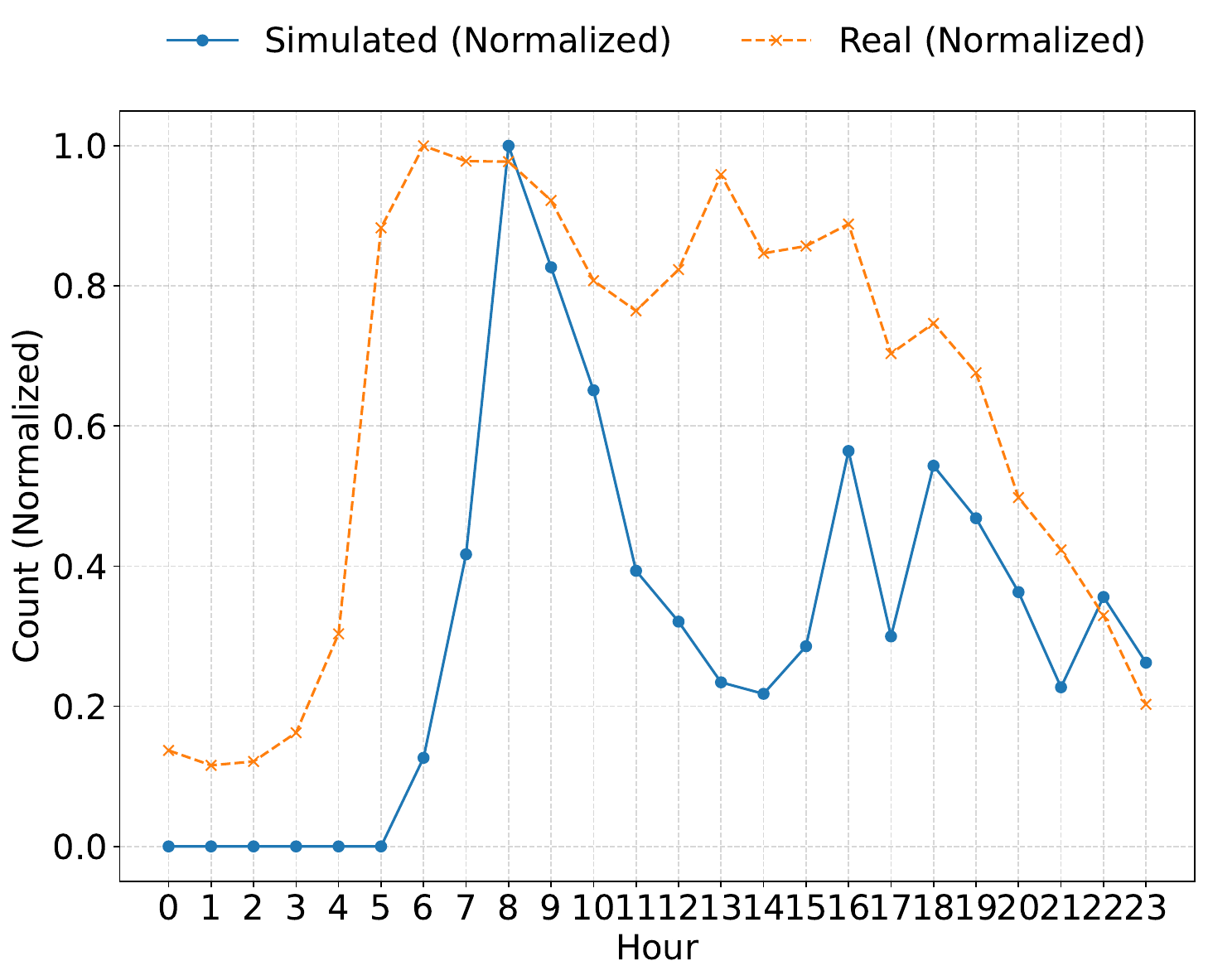}
	\caption[Traffic counts at TE038.]{Traffic counts from the traffic count station TE038.}
	\label{fig:1c_default:traffic_counts:TE038}
    \end{subfigure}
    \hfill
    \begin{subfigure}[b]{0.47\textwidth}
	\includegraphics[width=\linewidth]{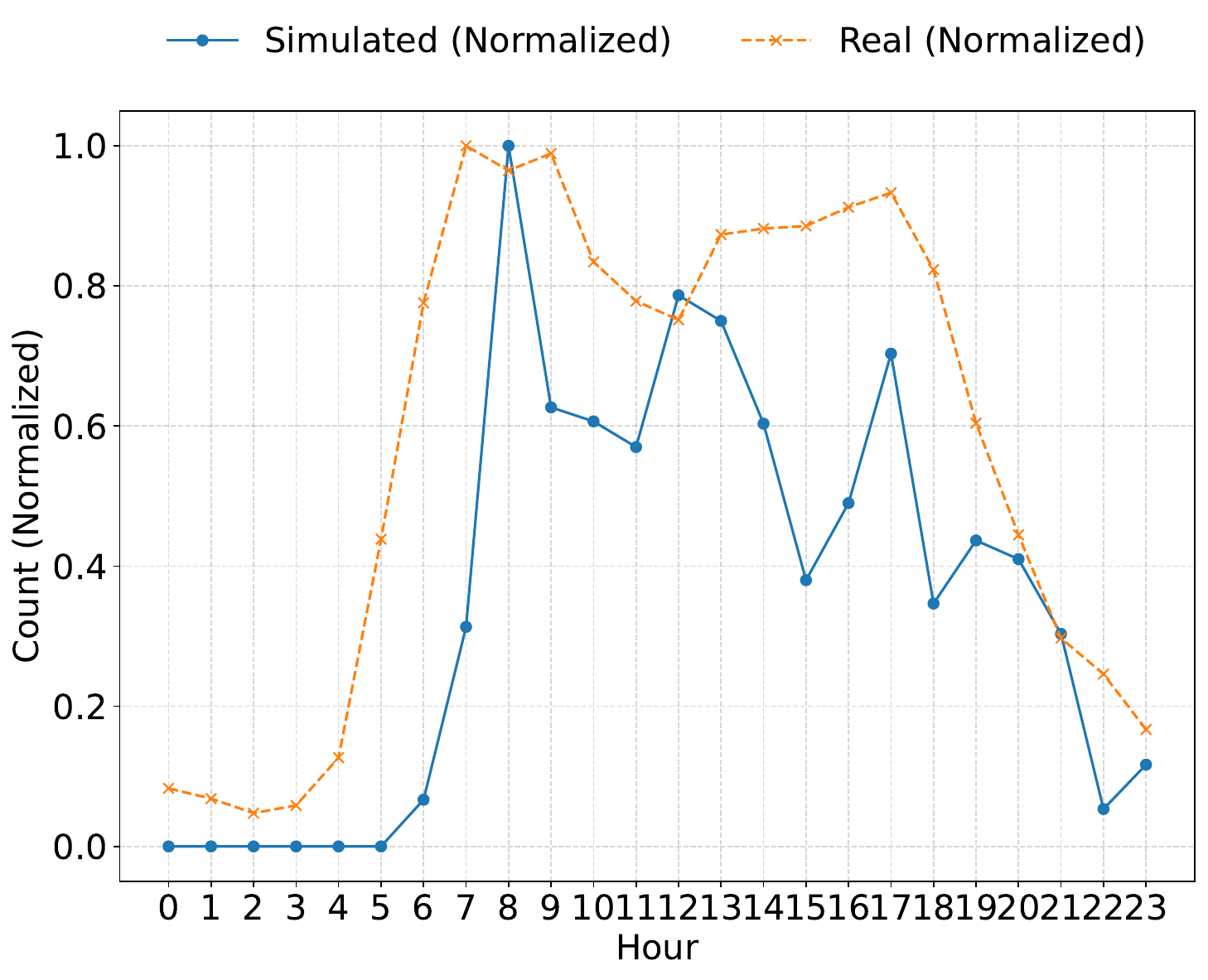}
	\caption[Traffic counts at TE042.]{Traffic counts from the traffic count station TE042.}
	\label{fig:1c_default:traffic_counts:TE042}
    \end{subfigure}
    \hfill
    \begin{subfigure}[b]{0.47\textwidth}
	\includegraphics[width=\linewidth]{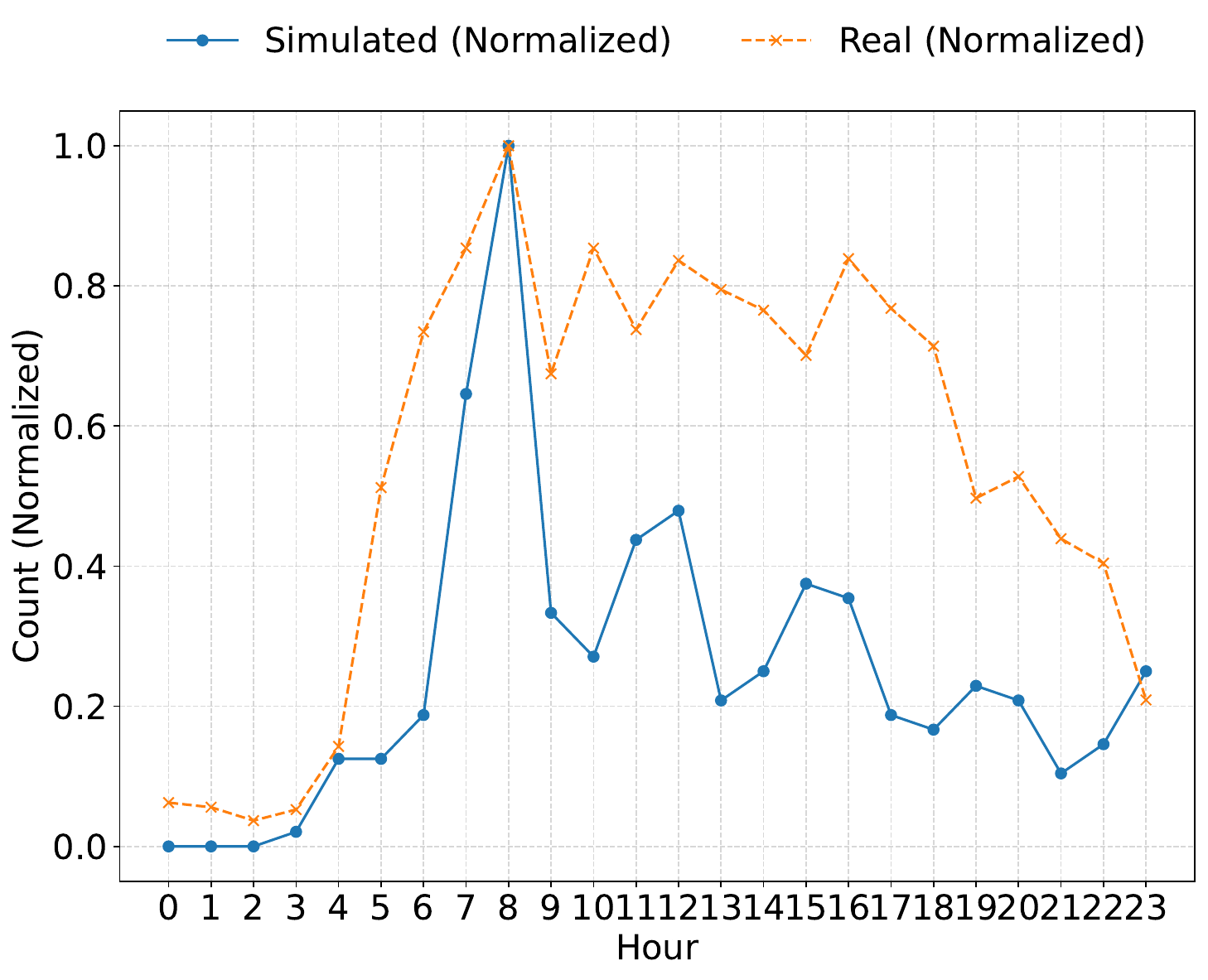}
	\caption[Traffic counts at TE244.]{Traffic counts from the traffic count station TE244.}
	\label{fig:1c_default:traffic_counts:TE244}
    \end{subfigure}
    \hfill
    \begin{subfigure}[b]{0.47\textwidth}
	\includegraphics[width=\linewidth]{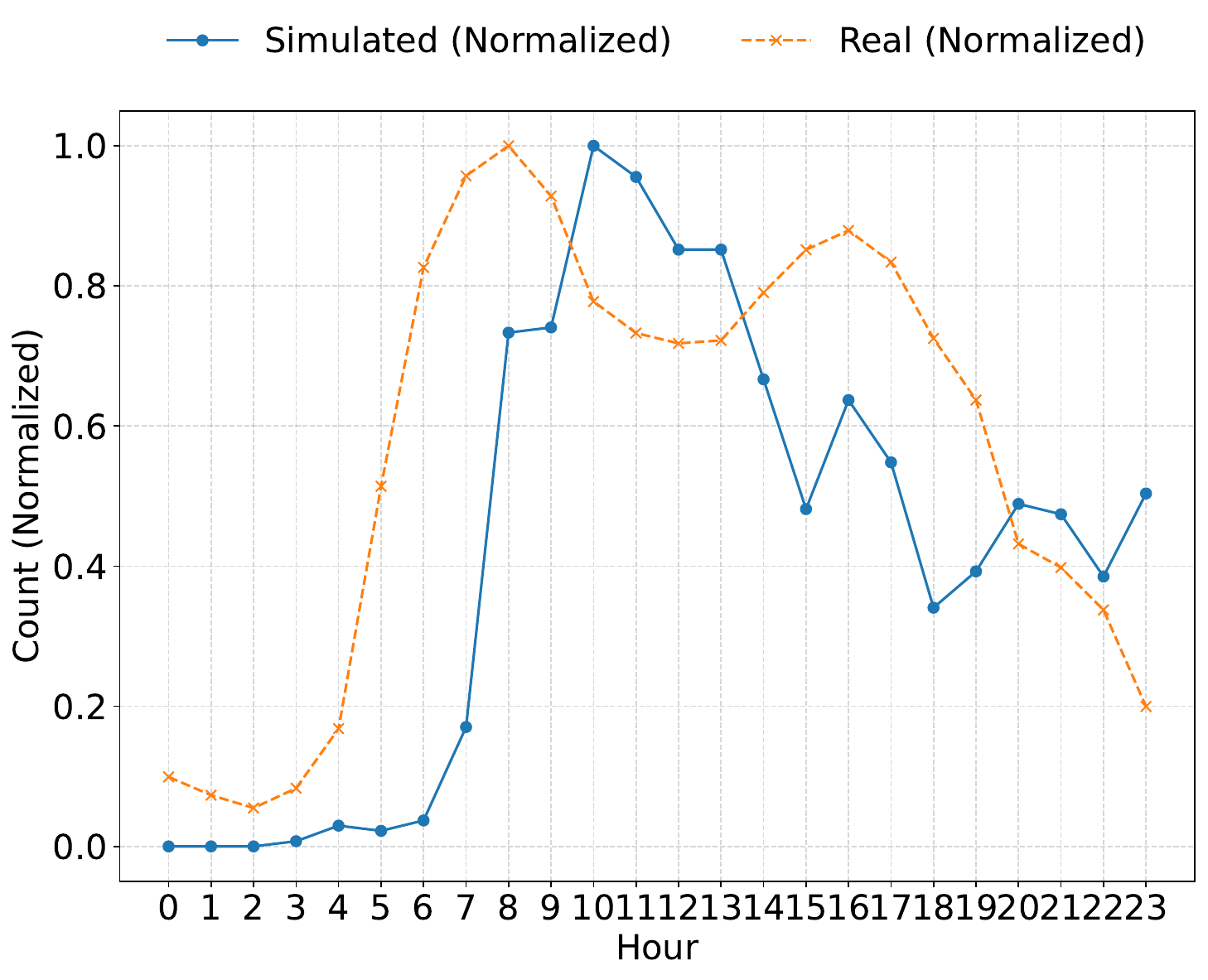}
	\caption[Traffic counts at TE309.]{Traffic counts from the traffic count station TE309.}
	\label{fig:1c_default:traffic_counts:TE309}
    \end{subfigure}
    \caption{Traffic counts at various traffic stations in the Wedding district.}
    \label{fig:traffic_counts}
    \Description{Four line charts compare hourly traffic counts from four stations between simulation and real-world data. Real counts remain high throughout the day, while simulated counts show sharper peaks in the morning.}
\end{figure*}

\subsection{Traffic Counts - \autoref{fig:traffic_counts}}

\ac{GTA} captures key temporal patterns in traffic flow, albeit with slightly lower overall volumes at the count stations.
Centrally located stations show higher simulated traffic than those Wedding's edges.
Nighttime traffic is nearly absent in \ac{GTA}, whereas real-world data still shows some activity.
Morning peaks are consistently reproduced but appear later and sharper than in reality. TE038 and TE042 both reflect midday dips, though shifted in time compared to actual counts. TE309 lacks a clear midday drop but shows a gradual decline, which may still reflect reduced midday activity. 
While replicating the morning peak, the traffic count at TE244 is lower for the remaining day. Overall, while \ac{GTA} slightly underestimates absolute volumes, it captures several realistic temporal trends across locations.

\section{Discussion}
This short paper introduces \textbf{population-grounded generative agents integrated with traffic
simulations} as a methodological bridge between survey-based modeling and early-stage innovation evaluation in mobility research.
By validating emergent behaviors against empirical data, we propose a scalable framework for designing realistic mobility behaviors. Unlike many recent agent-based approaches~\cite{park_generative_2023, park_social_2022} that focus on generating plausible individual behavior, \ac{GTA} goes further by producing behavior that is not only plausible at the individual level but also consistent with population-scale mobility patterns observed in empirical data. We demonstrate that \ac{GTA} captures the overall distribution of mode choices and models real-world trends in how economic status and occupation influence mobility choices.

\subsection{GTA as a Method to Simulate Situated Mobility at Scale}

\ac{GTA} bridges traditional agent-based models and persona-driven generative agents by combining census-aligned agent generation with \ac{LLM}-based contextual reasoning. Rather than relying on fixed utility functions, agents generate interpretable rationales for their mobility choices while remaining constrained by real-world networks.
A key contribution of \ac{GTA} is that this generative reasoning yields behavior consistent with empirical mobility patterns and observed traffic volumes. By evaluating GTA against mobility studies and traffic counts, \ac{GTA} moves beyond narrative plausibility toward behavioral realism. 
Demonstrating this alignment is essential because it validates natural-language reasoning as a reliable foundation for HCI simulations. 
We consider this a necessary foundation before generative agents can reliably support HCI applications.
For \ac{HCI}, \ac{GTA} provides a fast, low-cost method for prototyping mobility innovations. \ac{GTA} captures the situated interplay between socio-demographics, temporal context, and network structure while retaining natural-language interpretability, making it a practical basis for exploring mobility-related user experiences long before costly field studies are feasible.

\subsection{Behavioral Biases Toward ``Role Model Behavior''}
Our agents, regardless of socioeconomic profile, tend to exhibit behavior aligned with urban ideals: frequent biking, healthy life styles, and low nighttime activity. This homogeneity likely reflects the biases of both training data and model objectives, which prioritize plausibility over variability. LLMs often overrepresent socially desirable behaviors~\cite{rutinowski_self-perception_2024}, which might be due to their reliance on text~\citet{schroder_large_2025}. This ``role model paradox'' limits the diversity of agent behaviors.
A key challenge is introducing heterogeneity. This might not be achievable by varying personas alone, but by explicitly modeling behavioral priors, uncertainty, and human constraints in decision-making. Doing so could allow \ac{HCI} application to simulate diverse populations with greater fidelity.

\subsection{Trade-offs Between Rapid Prototyping and Empirical Validity}

Using \ac{LLM}-based agents as scalable proxies for human decision-making and cognition~\cite{binz_centaur_2024} has been criticized for introducing biases from textual training data, underrepresenting the variability of human behavior, and remaining fundamentally distinct from human minds~\cite{schroder_large_2025}. Our ablation experiments also reveal that the empirical outcome validity \ac{GTA} is highly sensitive to both prompting strategies and model choice (see \autoref{tab:rmse_summary}). Small changes, such as omitting CoT scaffolding or switching from Llama 3.1 to Qwen2.5, can significantly alter trip characteristics and distort mode choice distributions, even under identical conditions. A further challenge is ensuring behavioral consistency (e.g., an agent leaving work by car only if it arrived by car).

Nonetheless, \ac{GTA} is a valid method for rapid prototyping of mobility innovations.
We view \ac{LLM}-based simulations as tools for ideation, not replacements for empirical studies~\cite{hamalainen_evaluating_2023}. While validated here for Berlin, re-evaluation in other regions is necessary. Given its low cost compared to surveys, \ac{GTA} offers a practical means to augment current \ac{HCI} mobility methods~\cite{schmidt_simulating_2024}.

\subsection{Future Work on GTA}
Several extensions can increase \ac{GTA} utility.
First, traffic simulation remains a bottleneck: \ac{SUMO}’s equilibrium calculations are slow, and the traffic network requires manual improvements. Integrating the \ac{LLM}-driven action module with simulators like MATSim~\cite{matsim_reference_2016} could reduce manual preprocessing and support multi-day scenarios.  
Second, individual mobility behavior is partially influenced by
others~\cite{kim_social_2018, pike_conditional_2018}. \ac{GTA} could model simple social networks~\cite{park_generative_2023, park_social_2022}, allowing agents to
adapt activities or mode choices based on routines or invitations of connected agents.
While outside the scope of this short paper, a dedicated interactive front-end would make \ac{GTA} more accessible. 
Finally, to improve simulation accuracy, agents could incorporate multi-day memory and adaptation.
Persistent memory would allow capturing long-term behavioral shifts and policy impacts~\cite{liu2024llmagentbasedmodelingtransportationsystems} and could increase travel plan consistency.

\begin{acks}
The authors acknowledge the financial support by the Federal Ministry of Research, Technology and Space of Germany and by Sächsische Staatsministerium für Wissenschaft, Kultur und Tourismus in the programme Center of Excellence for AI-research „Center for Scalable Data Analytics and Artificial Intelligence Dresden/Leipzig“, project identification number: ScaDS.AI
\end{acks}

\section*{Open Science}
To support the open science movement in HCI~\cite{salehzadeh_niksirat_changes_2023, ebel_changing_2024}, all code, data, and analysis code are available here: \url{https://github.com/ciao-group/Generative-Traffic-Agents}


\pagebreak
\bibliographystyle{ACM-Reference-Format}
\bibliography{bib2, references}

\pagebreak
\appendix
\section{Appendix}\label{ch:appendix}

\begin{lstlisting}[style=distinct, caption={Prompt used to generate agent descriptions from sampled socio-demographic and socio-economic attributes.}, label={lst:description_prompt}, breaklines=true]
Sample attributes from the Berlin population: [sampled attributes]
Imagine a realistic person with these attributes. Write a specific one paragraph description:
[JSON prompt]
\end{lstlisting}

\begin{lstlisting}[style=distinct, caption={Prompt for generating daily activity plans.}, label={lst:daily_activity_prompt}, breaklines=true]
You are: [persona description]
Today is [date specification].
[optional information about external factors]
Write in broad strokes what you are doing during the day. Start the day at home.
Only include tasks that occur at a specific location which must be one of the provided building options and do not include any transportation or commuting tasks (for example, do not include actions like "walking by foot" or "driving a car" or "taking the bus") or locations (for example, "parking", bicycle_parking", etc.).
building options:
[relevant location categories]
[JSON prompt]
\end{lstlisting}

\begin{lstlisting}[style=distinct, caption={An excerpt of OSM attributes and their values.}, label={lst:exemplary_building_cat}, breaklines=true]
{
  'building': ['school', 'hospital', 'apartments', ...],
  'amenity': ['school', 'hospital', 'library', ...],
  'office': ['government', 'company', 'university', ...],
  'shop': ['supermarket', 'bakery', 'clothes', ...],
  'craft': ['carpenter', 'blacksmith', 'pottery', ...]
}
\end{lstlisting}

\begin{lstlisting}[style=distinct, caption={Prompt for means of transport decision.}, label={lst:transport_prompt}, breaklines=true]
You are: [persona description]

You live in Berlin and have several tasks to complete today, for which you need to plan several trips. Berliners typically
- walk for very short trips (<1 km), 
- bike for medium trips (1-5 km), 
- use public transport for longer trips (>5 km), 
- and drive only if they have a car immediately available.

The eight examples below are *shuffled*, but together they reflect a typical Berlin modal split (2 walk, 1 bicycle, 3 car, 2 public transport): 

1. 0.3 km in 5 min -> **walk** ("300 m is fastest on foot") 
2. 4.0 km in 10 min -> **car** ("fastest for a 4 km morning commute") 
3. 8.0 km in 20 min -> **public transportation** ("subway is most reliable") 
4. 1.5 km in 7 min -> **bicycle** ("Berlin's bike paths make this ideal") 
5. 0.7 km in 10 min (with groceries) -> **walk** ("easiest to carry bags") 
6. 5.0 km in 15 min -> **public transportation** ("smooth transfer on S-Bahn") 
7. 3.0 km in 12 min (rainy) -> **car** ("stay dry and faster than cycling") 
8. 12.0 km in 30 min -> **car** ("direct suburban route is best by car")

[optional information about external factors]
You live in Berlin. Today, you will go to several places. You have checked various ways to get from one place to another. For each route, first reason in one sentence how you would usually cover the distance, then select the means of transport. Switch transportation modes only when it logically fits the scenario. For example, you may only use your bicycle or car if it is already with you, and you should choose walking or public transit only if your bicycle or car remains at home or is scheduled to be picked up later. Your route options are as follows:
[possible route options]
[JSON prompt]
\end{lstlisting}



\begin{figure}[htbp]
  \centering
\includegraphics[width=\linewidth]{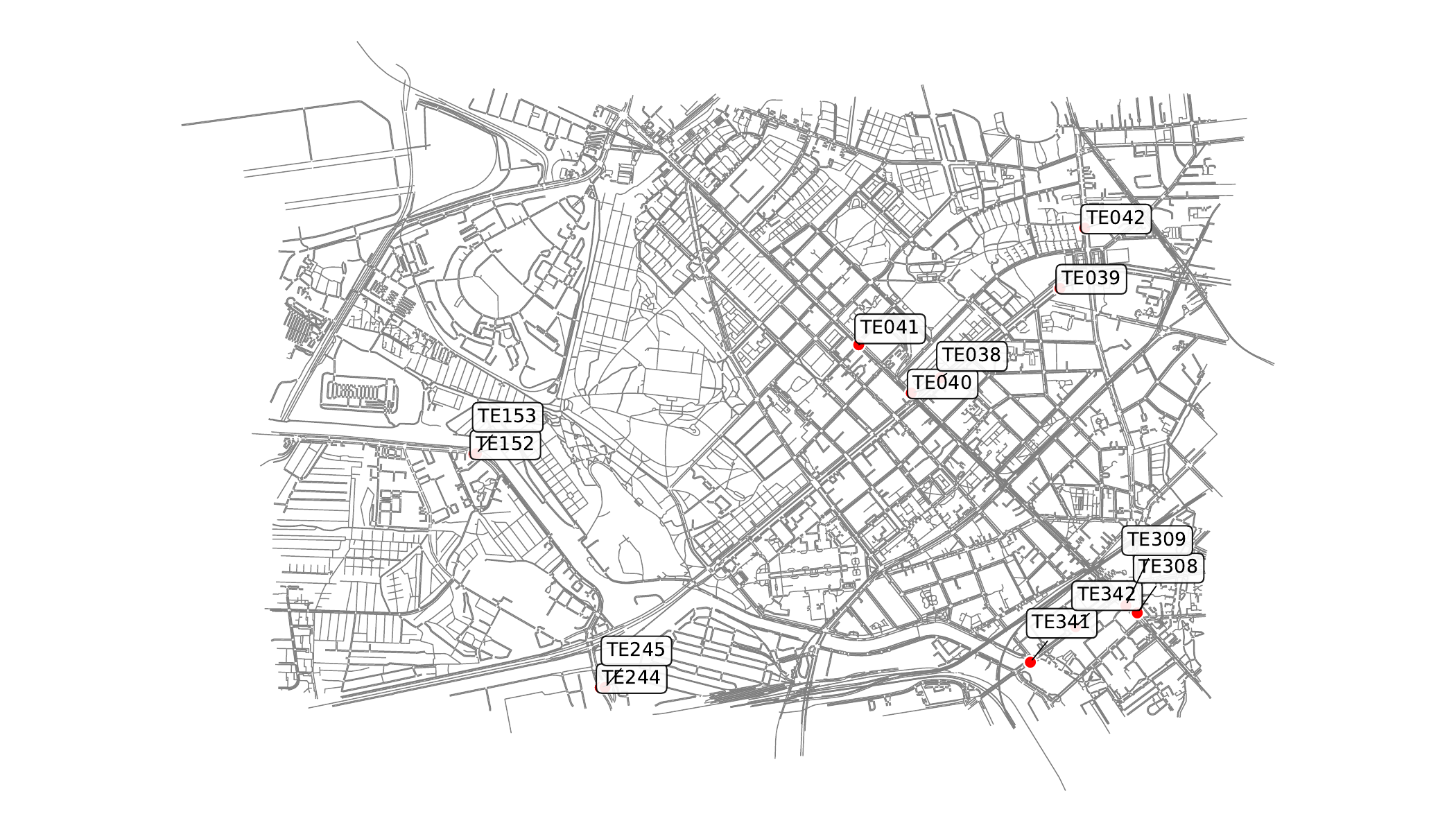}
  \caption{Map of traffic count stations in Berlin Wedding used for evaluation.}
  \label{fig:wedding_stations}
\end{figure}

\begin{lstlisting}[style=distinct, basicstyle=\scriptsize\ttfamily, caption={Description of Agent 3828.}, label={lst:transport_prompt2}, breaklines=true]
[DESCRIPTION]
(Meet 43-year-old Thomas, a married man living in Berlin with his wife and two children. As a full-time working professional, he spends most of his days commuting to the city center for work, although he occasionally uses his car for short trips around the city with his family. Thomas's household income is relatively stable, with a net monthly income of around 4,000-4,999 Euro, placing them in a high economic status. Despite not having a university degree, Thomas has adapted well to the demands of modern life in Berlin, often using public transportation and cycling with his family, although they do own one car for convenience. His family-oriented household consists of four members, and they enjoy a comfortable lifestyle in the city, taking advantage of the many amenities and opportunities Berlin has to offer.)

[DAY SCHEDULE]
 - 06:00: Waking up and getting ready for the day [house]
 - 07:00: Having breakfast with the family [house]
 - 08:00: Helping the children with their homework [house]
 - 09:00: Leaving for work after ensuring the family is prepared for the day [house]
 - 10:00: Attending a meeting with colleagues at the office [office]
 - 12:00: Having lunch at the office canteen [canteen]
 - 13:00: Continuing work at the office [office]
 - 17:00: Heading home from work [office]
 - 18:00: Spending quality time with the family, preparing dinner together [house]
 - 20:00: Eating dinner together as a family [house]
 - 21:00: Relaxing with the family in the evening [house]

[LOCATION CHANGES]
  Location Change 1:
    Purpose: Attending a meeting with colleagues at the office [office @ 10:00]
    From: house
    To:   office
    DECISION:
{       'means_of_transport': 'passenger',
        'reasoning': 'Given the short distance and time, using the car is the '
                     'fastest option for the morning commute to the office.',
        'route_id': 38280001}
  Location Change 2:
    Purpose: Having lunch at the office canteen [canteen @ 12:00]
    From: office
    To:   canteen
    DECISION:
{       'means_of_transport': 'pedestrian',
        'reasoning': 'As the distance is very short, walking is the most '
                     'convenient option to reach the office canteen.',
        'route_id': 38280102}
  Location Change 3:
    Purpose: Continuing work at the office [office @ 13:00]
    From: canteen
    To:   office
    DECISION:
{       'means_of_transport': 'pedestrian',
        'reasoning': "Although the distance is short, it's faster to walk to "
                     "the office, especially since we don't have to carry "
                     'groceries or other items.',
        'route_id': 38280203}
  Location Change 4:
    Purpose: Spending quality time with the family, preparing dinner together [house @ 18:00]
    From: office
    To:   house
    DECISION:
{       'means_of_transport': 'passenger',
        'reasoning': 'Given the moderate distance and the need to travel with '
                     'a family, using the car is the most practical option to '
                     'head home from work.',
        'route_id': 38280304}

\end{lstlisting}

\begin{lstlisting}[style=distinct, basicstyle=\scriptsize\ttfamily, caption={Description of Agent 1847.}, label={lst:transport_prompt3}, breaklines=true]
[DESCRIPTION]
Mrs. Mueller, a 78-year-old retiree, lives in a small household with her husband in a quiet Berlin suburb. She has a high level of education and enjoys a monthly net income of about 1,500 to 2,000 euros, which classifies her as having a low income. Mrs. Mueller is a typical representative of Berlin's senior population, having withdrawn into retirement and focusing on her hobbies and life with her partner. She owns a car, which she uses only occasionally, and has no need for a home office, as she is retired.

[DAY SCHEDULE]
 - 06:00: Waking up and having a morning coffee [house]
 - 07:00: Watching TV and checking the news [house]
 - 08:00: Preparing breakfast [house]
 - 09:00: Having breakfast with my husband [house]
 - 10:00: Going for a walk in our garden [house]
 - 11:00: Doing some light gardening [house]
 - 12:00: Having lunch [house]
 - 13:00: Watching a movie [house]
 - 15:00: Taking a short walk to the nearby park [house]
 - 16:00: Sitting on a bench in the park and socializing with neighbors [bench]
 - 18:00: Having dinner [house]
 - 19:00: Watching TV or reading a book [house]
 - 21:00: Going to bed [house]

[LOCATION CHANGES]

  Location Change 1:
    Purpose: Sitting on a bench in the park and socializing with neighbors [bench @ 16:00]
    From: house
    To:   bench
    DECISION:
{       'means_of_transport': 'pedestrian',
        'reasoning': "I'll walk to the nearby park as it's a short distance "
                     'and I can easily do it on foot',
        'route_id': 18470001}

  Location Change 2:
    Purpose: Having dinner [house @ 18:00]
    From: bench
    To:   house
    No decision.

\end{lstlisting}

\newpage

\begin{lstlisting}[style=distinct, basicstyle=\scriptsize\ttfamily, caption={Description of Agent 8001}, label={lst:transport_prompt4}, breaklines=true]

[DESCRIPTION]
('She is a 51-year-old woman living in Berlin with a moderate income, working full-time, and cycling to work as she has a household with a bike. She has a secondary education and is part of a household with three people, without a car. Her monthly household net income is between 2,000 and 3,000 euros, placing her in the middle socioeconomic class. She is likely to be an environmentally conscious and active member of her community, valuing independence and simplicity in her daily life.')

[DAY SCHEDULE]
 - 07:00: Having breakfast with my household members in our kitchen [house]
 - 08:00: Checking and responding to personal emails and messages before starting work [house]
 - 09:00: Leaving the house to get some fresh air and water the plants in our small garden [house]
 - 10:00: Working on a community project in our local community centre [community_centre]
 - 12:00: Meeting with colleagues for lunch at a nearby cafe [cafe]
 - 13:00: Attending a workshop on sustainable living at the local social centre [social_centre]
 - 16:00: Picking up some groceries at the nearby supermarket [supermarket]
 - 17:00: Preparing dinner with my household members in our kitchen [house]
 - 18:30: Spending time with family and friends at our house [house]
 - 20:00: Reviewing the day's activities and planning for the next day in our living room [house]

[LOCATION CHANGES]

  Location Change 1:
    Purpose: Working on a community project in our local community centre [community_centre @ 10:00]
    From: house
    To:   community_centre
    DECISION:
{       'means_of_transport': 'bicycle',
        'reasoning': 'Given the short distance and my preference for cycling, "I'll bike to the community centre.",
        'route_id': 80010001}
  Location Change 2:
    Purpose: Meeting with colleagues for lunch at a nearby cafe [cafe @ 12:00]
    From: community_centre
    To:   cafe
    DECISION:
{       'means_of_transport': 'public transport',
        'reasoning': "As I'll be meeting colleagues for lunch, I'll take the public transport for its reliability and slightly faster travel time.',
        'route_id': 80010102}
  Location Change 3:
    Purpose: Attending a workshop on sustainable living at the local social centre [social_centre @ 13:00]
    From: cafe
    To:   social_centre
    DECISION:
{       'means_of_transport': 'public transport',
        'reasoning': "Since I don't have a car available, I'll take public transport to the social centre.',
        'route_id': 80010203}
  Location Change 4:
    Purpose: Picking up some groceries at the nearby supermarket [supermarket @ 16:00]
    From: social_centre
    To:   supermarket
    DECISION:
{       'means_of_transport': 'pedestrian',
        'reasoning': 'Walking is the most convenient option for picking up groceries, allowing me to carry my bags easily.',
        'route_id': 80010304}
  Location Change 5:
    Purpose: Preparing dinner with my household members in our kitchen [house @ 17:00]
    From: supermarket
    To:   house
    DECISION:
{       'means_of_transport': 'pedestrian',
        'reasoning': "Given the very short distance, I'll walk back home with the groceries.',
        'route_id': 80010405}
\end{lstlisting}

\end{document}